\pdfoutput=1

\documentclass[12pt,nofootinbib]{article}
\usepackage{amsmath}
\usepackage{amssymb}
\usepackage{graphicx}
\usepackage[colorlinks=true,linkcolor=blue,citecolor=blue]{hyperref}

\numberwithin{equation}{section}

\newcommand{\GEV}{ {\rm GeV} }
\def\GEV#1{10^{#1}{\rm\,GeV}}

\begin{document}

\begin{titlepage}

\setcounter{page}{1} \baselineskip=15.5pt \thispagestyle{empty}

\begin{flushright}
{\footnotesize RESCEU-28/11}\\
{\footnotesize ICRR-Report-592-2011-9}\\
{\footnotesize IPMU11-0122}\\
{\footnotesize TU-885}
\end{flushright}
\vfil

\bigskip\
\begin{center}
{\LARGE \textbf{Non-Gaussianity from Curvatons Revisited}}
\vskip 15pt
\end{center}

\vspace{0.5cm}
\begin{center}
{\large Masahiro
 Kawasaki$^{\ast, \dagger}$\footnote{kawasaki@icrr.u-tokyo.ac.jp}, Takeshi
 Kobayashi$^{\ddagger}$\footnote{takeshi-kobayashi@resceu.s.u-tokyo.ac.jp},
 and Fuminobu Takahashi$^{\dagger, \star}$\footnote{fumi@tuhep.phys.tohoku.ac.jp}} 
\end{center}

\vspace{0.3cm}

\begin{center}
\textit{$^{\ast}$ Institute for Cosmic Ray Research, The University of
 Tokyo, \\ 5-1-5 Kashiwanoha, Kashiwa, Chiba 277-8582, Japan}\\

\vskip 4pt
\textit{$^{\dagger}$ Institute for the Physics and Mathematics of the
 Universe, The University of Tokyo, \\ 5-1-5 Kashiwanoha,
 Kashiwa, Chiba 277-8582, Japan}\\

\vskip 4pt
\textit{$^{\ddagger}$ Research Center for the Early Universe, School of
 Science, The University of Tokyo, \\ 7-3-1 Hongo, Bunkyo-ku, Tokyo
 113-0033, Japan}\\ 

\vskip 4pt
\textit{$^{\star}$ Department of Physics, Tohoku University, Sendai 980-8578, Japan}
\end{center} \vfil

\vspace{0.8cm}

\noindent
We investigate density perturbations sourced by a curvaton with a generic
 energy potential. The key feature of a curvaton potential which
 deviates from a quadratic is that the curvaton experiences a non-uniform
 onset of its oscillation. This sources additional contributions to the
 resulting density perturbations, and we especially find that the
 non-Gaussianity parameter~$f_{\mathrm{NL}}$ can become large with
 either sign no matter whether the curvaton dominates or subdominates the
 universe when it decays. Such non-quadratic curvaton potentials are
 required in order to produce a red-tilted density perturbation spectrum
 (without invoking large-field inflation), and are also motivated by
 explicit curvaton models based on microscopic physics. We further
 apply our generic results to the case where the curvaton is a
 pseudo-Nambu-Goldstone (NG) boson of a broken U(1) symmetry, and show
 that the resulting density perturbations are strongly enhanced towards
 the hilltop region of the energy potential, accompanied by a mild
 increase of the non-Gaussianity. Such hilltop NG curvatons can produce
 observationally suggested density perturbations under wide ranges of 
 inflation/reheating scales, and further predict the non-Gaussianity of
 the density perturbations to lie within the range $10 \lesssim
 f_{\mathrm{NL}} \lesssim 30$. 

\vfil

\end{titlepage}

\newpage
\tableofcontents

\section{Introduction}
\label{sec:intro}

The origin of the large-scale structures of our universe can be
explained by the presence of primordial density perturbations in the
early universe, which were of super-horizon scales and almost scale-invariant.
However, understanding how such density perturbations were seeded has
remained a big mystery for cosmology. 
On the other hand, cosmic
inflation~\cite{Starobinsky:1980te,Sato:1980yn,Guth:1980zm} has been
known to be able to generate super-horizon and (nearly) scale-invariant
fluctuations for scalar fields. 
(Early universe models that can generate such field fluctuations without
having an inflationary era have also been studied, see
e.g.~\cite{Mukohyama:2009gg,Rubakov:2009np,Creminelli:2010ba}.)
Then now, one may rephrase the question about the origin of the
large-scale structures as how such field fluctuations were converted into the
primordial density perturbations of the universe, which eventually lead
to structure formation. 

An attractive mechanism for creating the primordial density
perturbations from the field fluctuations is the curvaton
mechanism~\cite{Linde:1996gt,Enqvist:2001zp,Lyth:2001nq,Moroi:2001ct}, in which the
curvaton field possessing large-scale field fluctuations generates the
density perturbations while it oscillates about its potential minimum
and comes close to dominating the universe. The idea of the curvaton
mechanism is thus simple, and also has the merit that when embedded in
inflationary cosmology, it frees the inflaton from being responsible
for generating the perturbations and therefore drastically relaxes constraints
on inflationary model building. 
However, while there have been extensive works on the curvaton
mechanism, most of the literature has focused on curvatons 
possessing rather trivial potentials (e.g. quadratic potentials), thus
lead to common predictions for the density perturbations,
such as large/small non-Gaussianity in the perturbations depending on
whether the curvaton is subdominant/dominant upon decay.

In this paper, we explore density perturbations sourced by a curvaton
rolling along an arbitrary energy potential. 
The key feature of a curvaton potential which deviates from a quadratic
one is that the curvaton field with large-scale field fluctuations starts
its oscillation at different times at different patches of the
universe. 
Such behaviour contributes to the curvaton energy density perturbations
in addition to that sourced directly from the original field
fluctuations.
We find that such non-trivial conversion processes of
the field fluctuations into the density perturbations can lead to strong
enhancement/suppression of the density perturbations of the universe, as
well as for their non-Gaussianities. 
In particular, we will show that the non-Gaussianity
parameter~$f_{\mathrm{NL}}$ can become large with either sign no matter
if the curvaton dominates or subdominates the universe when it decays. 
Several previous works have considered specific cases, e.g., 
curvaton potentials with a non-quadratic polynomial term were studied
in~\cite{Enqvist:2005pg,Enqvist:2008gk,Enqvist:2009zf}, and cosine type potentials
in~\cite{Kawasaki:2008mc} which clearly pointed out that non-uniform onset of
curvaton oscillations can have important effects on the 
density perturbations. 
Related discussions can be found in~\cite{Sasaki:2006kq}, 
though such effects were not explicitly taken into account.
In this paper we carry out a systematic study of density
perturbations from curvatons with generic potentials through analytic
analyses. 

It is worthwhile to investigate curvaton potentials which deviate from
simple quadratic ones, both from observational and theoretical reasons:
Considering a curvaton whose field fluctuations were
generated during the inflationary era (which is the case studied in
this paper), in order for the curvaton to produce the density
perturbation spectrum which is red-tilted (i.e. with negative spectral
index~$n_s\! - \! 1$) as suggested by latest CMB 
observations~\cite{Larson:2010gs,Dunkley:2010ge,Hlozek:2011pc}
without relying on specific inflation mechanisms,\footnote{For a non-tachyonic
curvaton (i.e. a curvaton located along a potential with non-negative
curvature during inflation), the spectral index of the resulting density
perturbation spectrum is set by the time variation of the 
Hubble parameter during inflation as $ n_s -1 \geq 2 \dot{H}/H^2$
(cf.~(\ref{ns-1})). Hence, assuming single-field canonical slow-roll
inflation, the Lyth bound~\cite{Lyth:1996im} relates the spectral index
with the inflaton 
field~$\phi$ range as 
\begin{equation*}
 \frac{1}{M_p} \left| \frac{d\phi}{d\mathcal{N}} \right| \geq
 \sqrt{1-n_s},
\end{equation*}
where $M_p$ is the reduced Planck mass and $\mathcal{N}$ the e-folding number.
This shows that unless the curvaton is tachyonic during inflation, 
the WMAP~\cite{Larson:2010gs} central value $n_s  \approx 0.96$ 
requires a super-Planckian field range for the inflaton.\label{foot1}}
the curvaton needs to be located along a potential with negative
curvature during inflation (cf.~(\ref{ns-1})). This obviously
suggests the curvaton potential to take non-trivial forms.
Furthermore, explicit curvaton models constructed in the framework of
microscopic physics can naturally possess intricate energy 
potentials, e.g., curvatons in supersymmetric
models~\cite{Hamaguchi:2003dc}, a stringy curvaton model
in~\cite{Kobayashi:2009cm} from a D-brane
moving along the internal compactified space giving rise to a
multi-dimensional periodic curvaton potential.
\cite{Burgess:2010bz}~also discusses a string realization of the
curvaton, where the potential is a linear combination of exponential
terms.  

We also apply our general results to the case where the curvaton is a 
pseudo-Nambu-Goldstone (NG) boson~\cite{Dimopoulos:2003az} of a broken U(1)
symmetry. The periodic curvaton potential necessarily possesses not only
minima but also maxima, around which the potential curvature is negative so
that the resulting density perturbation spectrum can become
red-tilted. We explore the parameter space of NG curvatons producing a
red-tilted density perturbation spectrum as 
suggested by observations, and show that the allowed regions for the
inflationary energy scale and the reheating temperature (i.e. when the
inflaton decays) are strongly constrained, unless the NG curvaton is
located close to the maximum of its potential during inflation. In such
hilltop cases, 
the resulting density perturbations are extremely enhanced, thus relaxing
the bounds on the inflation/reheating scales. Non-Gaussianity of the
density perturbations also increases in the hilltop limit to values
$10 \lesssim f_{\mathrm{NL}} \lesssim 30$ in most part of the allowed 
parameter window, which will be tested by upcoming CMB
measurements. 
Such mild increase of non-Gaussianity towards the hilltop limit is
actually a rather generic feature of hilltop curvatons, as will be shown
in the paper. 

This paper is organized as follows. We derive analytic expressions
for density perturbations from curvatons with generic potentials in
Section~\ref{sec:pert}. Then in Section~\ref{sec:flat},
we move on to discuss special cases where the curvaton starts its
oscillation from a flat region of its potential, including hilltop
curvatons. It will be shown that in such cases, effects due to fluctuations
of the onset of the oscillation become crucial. As a simple
example of a curvaton model generating density perturbations that match
with observational constraints, we investigate
the case where the curvaton is a pseudo-NG boson in 
Section~\ref{sec:PNGC}. We present our conclusions in
Section~\ref{sec:conc}. Discussions on dynamics of scalar fields in an
expanding universe are provided in Appendix~\ref{app:SFD}.
While we focus on sinusoidal curvaton oscillations in the main body of
the paper, in Appendix~\ref{app:non-sinu} we further generalize our
discussions to incorporate non-sinusoidal curvaton oscillations. 
Curvatons along linear potentials are also discussed in
Appendix~\ref{app:non-sinu}.

\section{Density Perturbations from Curvatons}
\label{sec:pert}

In this paper we focus on
curvatons~\cite{Linde:1996gt,Enqvist:2001zp,Lyth:2001nq,Moroi:2001ct,Dimopoulos:2003ss}
whose field fluctuations are generated in the inflationary era (though
many features and results can be shared with curvatons with
non-inflationary fluctuation generating scenarios). Then a curvaton
should be a light field so that it acquires 
field fluctuations that are nearly scale-invariant and
Gaussian. Its energy density is considered to be initially negligible,
and after inflation the curvaton starts oscillating about its potential minimum
and behaves as nonrelativistic matter. As its energy density grows
relative to radiation sourced from the inflaton decay, the curvaton
fluctuations increasingly contribute to the density perturbations until
the curvaton decays or dominates the universe. 

One of the main points of this paper is that 
the original field fluctuations of the curvaton~$ \sigma$ during
inflation can give rise to perturbations of the curvaton energy density
through perturbing the time when the curvaton starts its oscillation,
as well as by directly perturbing its energy density~$\rho_{\sigma}$.
Let us schematically write down the resulting density perturbations of 
the universe as 
\begin{equation}
 \zeta \sim c_1 \frac{\delta \rho_{\sigma}}{\rho_{\sigma}} - 
c_2 \frac{\delta H_{\mathrm{osc}}}{H_{\mathrm{osc}}},
 \label{image}
\end{equation}
where $H_{\mathrm{osc}}$ is the Hubble parameter at the onset of the
curvaton oscillation. We have added a minus sign to the second term in
the right hand side to indicate that larger~$H_{\mathrm{osc}}$ gives an
earlier onset of the curvaton oscillation, which more quickly redshifts
away the curvaton energy density. 
(\ref{image}) illustrates that direct energy perturbations~$\delta
\rho_\sigma$ can be cancelled by effects from $\delta H_{\mathrm{osc}}$,
suppressing the linear order density perturbations with possible
enhancement of the non-Gaussianity. Moreover, when effects from $\delta
H_{\mathrm{osc}}$ become dominant over that from~$\delta \rho_\sigma$,
the resulting density perturbations become very different from cases
with the familiar quadratic curvaton potential. However we should also
stress that (\ref{image}) is rather schematic, and the actual
density perturbation spectrum is expressed as an involved combination of
the two effects, cf.~(\ref{NfuncX}). 

\vspace{\baselineskip}

In the following subsections, we explicitly compute the density
perturbations from curvatons with generic energy potentials. First
deriving an expression of the form similar to (\ref{image}) using the
$\delta\mathcal{N}$-formalism~\cite{Starobinsky:1986fxa,Sasaki:1995aw,Wands:2000dp,Lyth:2004gb}, 
we then analyze the curvaton dynamics and obtain analytic expressions
for the density perturbations. 

\begin{figure}[tbp]
\begin{center}
  \begin{center}
 \includegraphics[width=.65\linewidth]{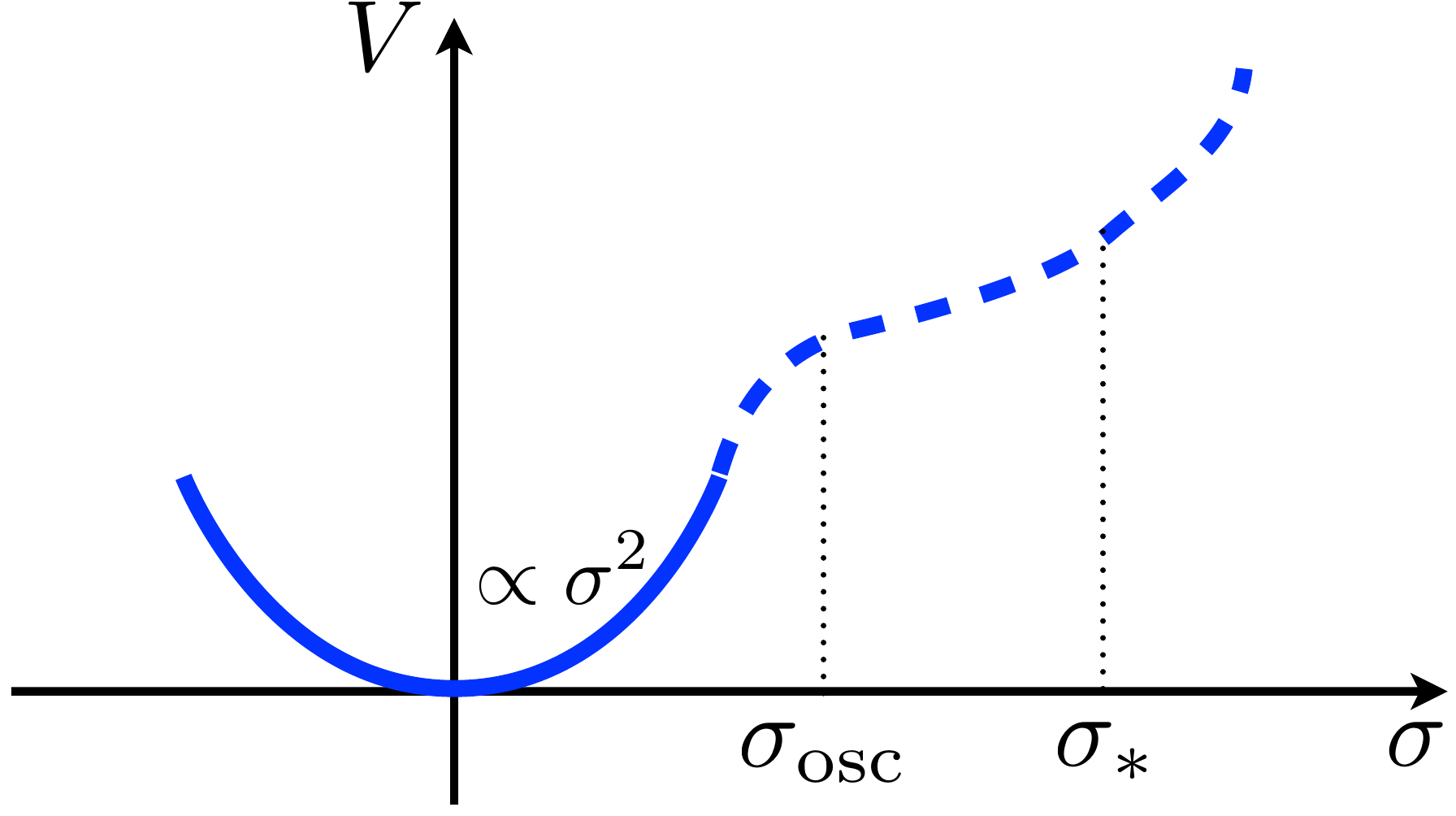}
  \end{center}
  \caption{Schematic of the curvaton potential under consideration:
 The curvaton rolls along its energy potential following an attractor
 solution~(\ref{eq17}), then it starts oscillating about the (local) minimum. The
 potential can take arbitrary forms, but once the curvaton
 starts the oscillation (from~$\sigma_{\mathrm{osc}}$) its amplitude
 soon decays and the potential is well approximated by a quadratic one.}
  \label{fig:schematic}
\end{center}
\end{figure}

We suppose that the Hubble parameter~$H = \dot{a}/a$ (with $a$
being the scale factor of the universe, and an overdot denoting a time
derivative) is nearly constant during inflation, and that 
at the end of inflation, the inflaton~$\phi$ suddenly turns into
matter, i.e. its energy density starts decreasing as $\rho_\phi
\propto a^{-3}$, and then at reheating, suddenly decays into
radiation, i.e. starts decreasing as $\rho_{r} \propto
a^{-4}$. Curvature perturbations sourced from the inflaton are ignored.
Furthermore, upon carrying out analytic estimations, 
we make the following simplifying assumptions for the curvaton:
\begin{enumerate}
 \item The curvaton~$\sigma$ starts a sinusoidal oscillation at some
       time after inflation, from then we treat the curvaton as
       matter. (Hence we are considering cases where the curvaton
       potential is well approximated by quadratic close to its (local)
       minimum, and that the curvaton oscillation happens mostly in the
       quadratic regime, cf. Figure~\ref{fig:schematic}. However, one
       can straightforwardly generalize 
       the calculations to other cases, as is discussed in
       Appendix~\ref{app:non-sinu}.) \label{ass1} 
 \item The energy density of the curvaton until it starts oscillating is
       so tiny compared to the total energy density that its effect on
       the expansion of the universe is negligible at least until the
       onset of the curvaton oscillation. \label{ass4}
\end{enumerate}
The curvaton acting as matter is also assumed to suddenly decay into
radiation when $H = \Gamma_\sigma$, where we take the curvaton decay
rate~$\Gamma_\sigma$ to be a uniform constant independent of the
curvaton field value.

\subsection{Computing $\delta \mathcal{N}$}
\label{subsec:deltaN}

The curvature perturbations~$\zeta$ from a curvaton can be computed using the 
$\delta\mathcal{N}$-formalism~\cite{Starobinsky:1986fxa,Sasaki:1995aw,Wands:2000dp,Lyth:2004gb}, 
as the difference in the e-folding number~$\mathcal{N}$ at a
constant energy hypersurface,
among different patches of the universe which took different field values
of the curvaton at the time when the CMB scale exited the horizon during
inflation, i.e.,
\begin{equation}
 \zeta (\boldsymbol{x}) = \frac{\partial \mathcal{N}}{\partial \sigma_*}
 \delta \sigma_*
  (\boldsymbol{x}) + \frac{1}{2} 
 \frac{\partial^2 \mathcal{N}}{\partial \sigma_*^2} 
 \left(
 \delta  \sigma_*(\boldsymbol{x})^2 
  - \langle \delta  \sigma_*(\boldsymbol{x})^2  \rangle
 \right)+ \cdots .
 \label{deltaN}
\end{equation}
Here, we are assuming a homogeneous and isotropic background, and 
values at the horizon exit are denoted by the subscript~$*$.
Expanding $\zeta(\boldsymbol{x})$ as
\begin{equation}
 \zeta_{\boldsymbol{k}} = \int d^3 \boldsymbol{x}\, 
  e^{-i \boldsymbol{k   \cdot x}} \zeta(\boldsymbol{x}),
\end{equation}
their two- and three-point correlation functions take the form
\begin{equation}
  \langle \zeta_{\boldsymbol{k_1}} \zeta_{\boldsymbol{k_2}} \rangle =
 (2 \pi)^3  \delta^{(3)} (\boldsymbol{k_1} + \boldsymbol{k_2}) P_\zeta (k_1),
\end{equation}
\begin{equation}
  \langle \zeta_{\boldsymbol{k_1}} \zeta_{\boldsymbol{k_2}}
   \zeta_{\boldsymbol{k_3}} \rangle = 
 (2 \pi)^3  \delta^{(3)} (\boldsymbol{k_1} + \boldsymbol{k_2} + \boldsymbol{k_3})
B_{\zeta} (k_1, k_2, k_3),
\end{equation}
where $k_i \equiv |\boldsymbol{k_i}|$.
We define the power spectrum of the density
perturbations~$\mathcal{P}_\zeta$ as 
\begin{equation}
 P_\zeta (k) = \frac{2 \pi^2}{k^3} \mathcal{P}_{\zeta} (k),
\label{power}
\end{equation}
and also the non-linearity
parameter~$f_{\mathrm{NL}}$~\cite{Komatsu:2001rj} as
\begin{equation}
 B_\zeta (k_1, k_2, k_3) = \frac{6}{5} f_{\mathrm{NL}}(k_1, k_2, k_3)
 \left[
 P_\zeta (k_1) P_\zeta (k_2) + P_\zeta (k_2) P_\zeta (k_3) + P_\zeta
 (k_3) P_\zeta (k_1) 
 \right],
\end{equation}
which parametrizes the non-Gaussianity of the density perturbations
encoded in the three-point function. 
From~(\ref{deltaN}), these values can be evaluated in terms of the derivatives
of~$\mathcal{N}$, at the leading order in $\delta \sigma_{*
\boldsymbol{k}}$, as
\begin{equation}
 \mathcal{P}_\zeta  = \left(\frac{ \partial \mathcal{N}}{\partial
		       \sigma_*}  \right)^2 
 \mathcal{P}_{\delta \sigma_*},
\end{equation}
\begin{equation}
 f_{\mathrm{NL}}  = 
 \frac{5}{6} \frac{\partial^2 \mathcal{N}}{\partial
  \sigma_*^2} 
 \left( \frac{\partial \mathcal{N}}{\partial \sigma_*} \right)^{-2},
\end{equation}
where $\mathcal{P}_{\delta\sigma_*}$ denotes the power spectrum of the
curvaton field fluctuations defined similarly as in (\ref{power}). Moreover,
we have considered the intrinsic non-Gaussianity of the field fluctuations
to be negligibly small and thus ignored the three-point function
for~$\delta\sigma_*$.

\subsubsection{Case $t_{\mathrm{reh}} < t_{\mathrm{osc}}$}

Let us begin by studying the case where reheating (= inflaton decay, at
$t_{\mathrm{reh}}$) happens before the onset of the curvaton oscillation
at $t_{\mathrm{osc}}$. In other words, the universe is radiation 
dominated when the curvaton starts oscillating. 
(Hereafter, values at time~$t_i$ ($i = $ reh, osc, etc.) are
denoted with the same subscript.)
We would like to derive the $\sigma_*$-dependence of the e-folding
number 
\begin{equation}
 \mathcal{N} = \mathcal{N}_a + \mathcal{N}_b,
\end{equation}
where
\begin{equation}
 \mathcal{N}_a \equiv \int^{t_{\mathrm{osc}}}_{t_*} H(t') dt', \qquad
 \mathcal{N}_b \equiv  \int^{t_{\mathrm{dec}}}_{t_{\mathrm{osc}}} H(t') dt'.
\end{equation}
Here $t_{\mathrm{dec}}$ denotes the time when the curvaton decays,
i.e. when $H$ becomes equal to $\Gamma_\sigma$ (which is independent of
$\sigma_*$). After~$t_{\mathrm{dec}}$,
the universe is filled with a single component,
i.e. radiation, hence no extra $\delta \mathcal{N}$ is produced. 

The energy density of radiation redshifts away as 
$\dot{\rho_r} = -4 H \rho_r$, thus 
\begin{equation}
 \mathcal{N}_a = 
  \int_{t_*}^{t_{\mathrm{reh}}} H(t') dt'
 +  \int^{t_{\mathrm{osc}}}_{t_{\mathrm{reh}}} H(t') dt'
 = \mathrm{const.} + \frac{1}{4} \ln 
 \frac{\rho_{r \mathrm{reh}}}{\rho_{r \mathrm{osc}}},
\end{equation}
where we have especially used Assumption~\ref{ass4}. Since $3 M_p^2
H_{\mathrm{osc}}^2 = \rho_{r\mathrm{osc}}$ (where $M_p$ is the reduced
Planck mass), we get
\begin{equation}
 \frac{\partial \mathcal{N}_a}{\partial \sigma_*} = 
 - \frac{1}{4} \frac{\partial}{\partial \sigma_*} 
 \ln H_{\mathrm{osc}}^2. \label{Na}
\end{equation}
Similarly, one obtains
\begin{equation}
 \mathcal{N}_b = 
 \frac{1}{4} \ln 
 \frac{\rho_{r \mathrm{osc}}}{\rho_{r \mathrm{dec}}}
 = 
 \frac{1}{4} \ln 
 \frac{3 M_p^2 H_{\mathrm{osc}}^2}{ 3 M_p^2 \Gamma_\sigma^2 -
 \rho_{\sigma \mathrm{dec}}} ,
\end{equation}
and by further using 
\begin{equation}
 \rho_{\sigma \mathrm{dec}} = \rho_{\sigma
\mathrm{osc}} e^{-3 \mathcal{N}_b},
\end{equation}
one can check that
\begin{equation}
 \frac{\partial \mathcal{N}_b}{\partial \sigma_*} = \frac{1}{4+3 r} 
  \left( 
  r \frac{\partial}{\partial \sigma_*}  \ln \rho_{\sigma\mathrm{osc}}
 + \frac{\partial}{\partial \sigma_*} 
 \ln H_{\mathrm{osc}}^2
 \right).  \label{Nb}
\end{equation}
Here we have introduced
\begin{equation}
 r \equiv \left. \frac{\rho_{\sigma}}{\rho_r} \right|_{\mathrm{dec}}.
 \label{efr}
\end{equation}
Therefore, from (\ref{Na}) and (\ref{Nb}) we arrive at 
\begin{equation}
 \frac{\partial \mathcal{N}}{ \partial \sigma_*} = 
 \frac{r}{4+3 r} \frac{\partial}{\partial \sigma_*} 
 \left( \ln \rho_{\sigma \mathrm{osc}} - \frac{3}{4} \ln
 H_{\mathrm{osc}}^2 \right), \label{AdelN}
\end{equation}
which reproduces the schematic expression of (\ref{image}). 
We note that the second term in the right hand side that is due to
the non-uniform onset of the curvaton oscillations, can become important
for non-quadratic curvaton potentials.\footnote{Strictly speaking,
the first term also contains effects due to non-uniform onset of the
oscillations through $\partial \sigma_{\mathrm{osc}} / \partial
\sigma_*$, as we will see later. 
Effects from~$\delta H_{\mathrm{osc}}$ displayed in~(\ref{AdelN}) 
have not been taken into account explicitly in previous works giving
general discussions on density perturbations in the curvaton scenario.}

Computation of the second derivative of~$\mathcal{N}$ is also straightforward. 
Since
\begin{equation}
 r = \left. \frac{\rho_{\sigma}}{\rho_{r}} \right|_{\mathrm{osc}}  e^{\mathcal{N}_b},
\end{equation}
one finds
\begin{equation}
 \frac{\partial r}{ \partial \sigma_*} = 4 (1+r) \frac{\partial
  \mathcal{N}}{\partial \sigma_*} , \label{Ar}
\end{equation}
thus giving 
\begin{equation}
 \frac{\partial^2 \mathcal{N}}{ \partial \sigma_*^2} = 
 \frac{16 (1+r)}{ (4+3r) r} 
 \left( \frac{\partial \mathcal{N}}{ \partial \sigma_*}\right)^2
 +  \frac{r}{4+3 r} \frac{\partial^2}{\partial \sigma_*^2} 
 \left( \ln \rho_{\sigma \mathrm{osc}} - \frac{3}{4} \ln
 H_{\mathrm{osc}}^2 \right). \label{ANN}
\end{equation}

\subsubsection{Case $t_{\mathrm{reh}} > t_{\mathrm{osc}}$}

The case of $t_{\mathrm{reh}} > t_{\mathrm{osc}}$ can also be computed
in a similar fashion as above. Since the curvaton
energy density is assumed to be negligibly small compared to the total
energy at the onset of the oscillation, the curvaton energy
continues to be negligible until reheating. 

The linear order perturbation is
\begin{equation}
 \frac{\partial \mathcal{N}}{ \partial \sigma_*} = 
 \frac{r}{4+3 r} \frac{\partial}{\partial \sigma_*} 
 \left( \ln \rho_{\sigma \mathrm{osc}} - \ln
 H_{\mathrm{osc}}^2 \right),
\end{equation}
where the coefficient of $\ln H_{\mathrm{osc}}^2$ differs from
(\ref{AdelN}) due to different expansions of the
universe (i.e. radiation or matter dominated) when the curvaton starts
to oscillate. 

(\ref{Ar}) remains the same, hence the second derivative of
$\mathcal{N}$ is
\begin{equation}
 \frac{\partial^2 \mathcal{N}}{ \partial \sigma_*^2} = 
 \frac{16 (1+r)}{ (4+3r) r} 
 \left( \frac{\partial \mathcal{N}}{ \partial \sigma_*}\right)^2
 +  \frac{r}{4+3 r} \frac{\partial^2}{\partial \sigma_*^2} 
 \left( \ln \rho_{\sigma \mathrm{osc}} -  \ln
 H_{\mathrm{osc}}^2 \right). 
\end{equation}

\subsection{Curvaton Field Value at the Onset of Oscillation}
\label{subsec:CFVOC}

In order to evaluate the expressions derived above, we need to know the
oscillation amplitude of the curvaton at a certain time, e.g., the
curvaton field value at the onset of oscillation~$\sigma_{\mathrm{osc}}$.
Considering a curvaton rolling along an energy potential~$V$
which is a function only of $\sigma$ (hence we do not consider
time-dependent potentials, though extending the discussion to such
potentials is straightforward), then the curvaton dynamics before it
starts its oscillation can be tracked by the attractor solution
\begin{equation}
 c H \dot{\sigma} \simeq -V', \label{eq17}
\end{equation}
where a prime denotes a derivative with respect to~$\sigma$, 
and $c$ is a constant taking the following values depending on the
expansion of the universe:
\begin{equation}
 c = 
 \left\{
   \begin{array}{cl}
     3 & \mbox{(de~Sitter) } \\
     9/2 & \mbox{(matter domination) } \\
     5 & \mbox{(radiation domination) } 
   \end{array}
\right.
 \label{ciroiro}
\end{equation}
Detailed discussions on this approximation are provided in Appendix~\ref{app:SFD}. 
Here we assume that the approximation~(\ref{eq17}) holds until $t_{\mathrm{osc}}$,
when the curvaton suddenly starts a sinusoidal oscillation,
cf. Figure~\ref{fig:schematic}. 
Validity of this assumption will be discussed in the next subsection.

Integrating
\begin{equation}
 \frac{d\sigma}{V'} = -\frac{dt}{c H}
\end{equation}
from the time of horizon exit  until the onset of the oscillation, one
obtains
\begin{equation}
 \int^{\sigma_{\mathrm{osc}}}_{\sigma_*} \frac{d\sigma}{V'} = \mathrm{const.} -
  \int^{H_{\mathrm{osc}}} \frac{dH}{c H \dot{H}}, \label{19}
\end{equation}
where $c$ in the right hand side is the one right before $t_{\mathrm{osc}}$,
taking either 5 or $ 9/2$ depending on whether $t_{\mathrm{reh}} <
t_{\mathrm{osc}}$ or $t_{\mathrm{reh}} > t_{\mathrm{osc}}$,
respectively.\footnote{Strictly speaking, here we are further assuming
that $c$ stays constant 
around~$t_{\mathrm{osc}}$, so that $c$ is independent of the curvaton
field fluctuations. So, e.g., the time of reheating should not be too
close to $t_{\mathrm{osc}}$.} We have made use of
Assumption~\ref{ass4} upon ignoring the $\sigma_*$-dependence of the
other terms denoted as $\mathrm{const.}$

Supposing that $H_{\mathrm{osc}}$ is determined merely by
$\sigma_{\mathrm{osc}}$, i.e. $H_{\mathrm{osc}} = H_{\mathrm{osc}}
(\sigma_{\mathrm{osc}})$ (cf.~(\ref{2424})), then after differentiating both sides 
of (\ref{19}) by~$\sigma_*$, one finds
\begin{equation}
 \frac{\partial \sigma_{\mathrm{osc}}}{\partial \sigma_*} = 
 \left\{1 -
  \frac{1}{c (c-3)}
\frac{V'(\sigma_{\mathrm{osc}})}{H_{\mathrm{osc}}^3}
 \frac{\partial H_{\mathrm{osc}}}{\partial \sigma_{\mathrm{osc}}}
 \right\}^{-1}
 \frac{V'(\sigma_{\mathrm{osc}})}{V'(\sigma_*)}. \label{sigmaoscstar}
\end{equation}

The value of $\sigma_{\mathrm{osc}}$ itself can, in principle, also be computed
from~(\ref{19}). Taking into account that 
$H \simeq H_{\mathrm{inf}} = \mathrm{const.}$ and 
$c=3$ during inflation, and that $\dot{H}/H^2 = 3-c$ for 
matter/radiation domination, one obtains
for $t_{\mathrm{reh}} < t_{\mathrm{osc}}$:
\begin{equation}
  \int^{\sigma_{\mathrm{osc}}}_{\sigma_*} \frac{d\sigma}{V'} = 
 -\frac{\mathcal{N}_*}{3 H_{\mathrm{inf}}^2} 
 + \frac{2}{27} \left(\frac{1}{H_{\mathrm{inf}}^2} - \frac{1}{
		 H_{\mathrm{reh}}^2}\right) 
 + \frac{1}{20} \left(\frac{1}{H_{\mathrm{reh}}^2} - \frac{1}{
		 H_{\mathrm{osc}}^2}\right),
\end{equation}
and for $t_{\mathrm{reh}} > t_{\mathrm{osc}}$:
\begin{equation}
  \int^{\sigma_{\mathrm{osc}}}_{\sigma_*} \frac{d\sigma}{V'} = 
 -\frac{\mathcal{N}_*}{3 H_{\mathrm{inf}}^2} 
 + \frac{2}{27} \left(\frac{1}{H_{\mathrm{inf}}^2} - \frac{1}{
		 H_{\mathrm{osc}}^2}\right),
\end{equation}
where $\mathcal{N}_*$ is the number of e-foldings during inflation
between the horizon exit of the CMB scale and the end of
inflation. Hence for $H_{\mathrm{inf}}^2 \gg H_{\mathrm{osc}}^2 $ (and
further $H_{\mathrm{reh}}^2 \gg H_{\mathrm{osc}}^2 $ for
$t_{\mathrm{reh}} < t_{\mathrm{osc}}$), we get
\begin{equation}
 \int^{\sigma_{\mathrm{osc}}}_{\sigma_*} \frac{d\sigma}{V'} \simeq 
 -\frac{\mathcal{N}_*}{3 H_{\mathrm{inf}}^2} - \frac{1}{2 c (c-3)
 H_{\mathrm{osc}}^2},  \label{sigmaosc}
\end{equation}
where $c$ is the value right before $t_{\mathrm{osc}}$. The two terms in
the right hand side can be comparable, e.g., when setting the curvaton
(effective) mass squared as $m^2 \sim H_{\mathrm{inf}}^2 / 100 $ so that
the spectral index becomes $|n_s - 1 |\sim 10^{-2}$, then 
$H_{\mathrm{osc}}^2 \sim m^2 \sim H_{\mathrm{inf}}^2 / 100 \sim
H_{\mathrm{inf}}^2 / \mathcal{N}_*$. 
Since $H_{\mathrm{osc}}^2$ can be given by the curvaton potential as we
will soon see, one can derive $\sigma_{\mathrm{osc}}$ as a function
of $\sigma_*$ by solving (\ref{sigmaosc}).

\subsection{When the Curvaton Starts Oscillating}
\label{sec:H_osc}

The last piece of information needed to evaluate the density
perturbation spectrum is when the curvaton starts its oscillation,
i.e.,~$H_{\mathrm{osc}}$. From the necessary
condition~(\ref{sscond}) for the attractor solution~(\ref{eq17}) with
(\ref{thisisc}) to hold, one may guess that the oscillation starts when
$H^2 = |V''|/c$. 
However actually, the curvaton velocity does not increase suddenly at 
$H^2 = |V''|/c$, but the curvaton stays on the approximation of
the form $H \dot{\sigma} \propto -V'$ for some more time. (See also
later discussions in~Subsection~\ref{sec:val}.)
Moreover, the curvaton being on the attractor~(\ref{eq17}) does not
necessarily imply that the curvaton oscillation has not yet started. 

Here, instead, we define the onset of the oscillation as when
the time scale of the curvaton rolling becomes comparable to
the Hubble time, i.e.
\begin{equation}
  \left| \frac{\dot{\sigma}}{H \sigma} \right|_{\mathrm{osc}} 
\sim 1, \label{dotsigma}
\end{equation}
where we have assumed that the (local) minimum of the curvaton potential
is set to $\sigma = 0$.
Here it should be noted that modifying (\ref{dotsigma}) by constant
factors (e.g., choosing the right hand side to be, say, 2) in many cases
results merely in modifying~$H_{\mathrm{osc}}$ by constant factors,
whose contributions to the density perturbations through $\delta \ln
H_{\mathrm{osc}}$ drop out, cf.~(\ref{AdelN}). Hence throughout this
paper we fix the right hand side of the condition~(\ref{dotsigma}) to be
unity. 

Then, supposing that the approximation (\ref{eq17}) holds until~$t_{\mathrm{osc}}$
(which is a valid assumption in many cases since, as mentioned above,
approximation of the form~(\ref{eq17}) does not
break down suddenly at $H^2 = |V''|/c$), one can 
rewrite (\ref{dotsigma}) as
\begin{equation}
 H_{\mathrm{osc}}^2 = 
 \left| \frac{V'(\sigma_{\mathrm{osc}})}{c \sigma_{\mathrm{osc}}}
 \right|
 = \frac{V'(\sigma_{\mathrm{osc}})}{c \sigma_{\mathrm{osc}}}. \label{2424}
\end{equation}
Here $c$ is the value right before~$t_{\mathrm{osc}}$, and we have
removed the absolute value sign in the far right hand side by assuming that the 
curvaton potential is monotonically increasing (decreasing) for $\sigma
> (<) 0$ (otherwise the curvaton following the attractor~(\ref{eq17})
stops rolling before reaching the origin). Thus we obtain 
\begin{equation}
 \frac{1}{H_{\mathrm{osc}}}\frac{\partial H_{\mathrm{osc}}}{\partial
  \sigma_{\mathrm{osc}}} = \frac{1}{2}
 \left(\frac{V''(\sigma_{\mathrm{osc}})}{V'( \sigma_{\mathrm{osc}} )} -
  \frac{1}{\sigma_{\mathrm{osc}}}  \right).
\end{equation}
Together with (\ref{sigmaoscstar}) this yields
\begin{equation}
\frac{\partial \sigma_{\mathrm{osc}}}{\partial \sigma_*} = 
\left\{
1 - \frac{1}{2 (c-3)} \left(\frac{\sigma_{\mathrm{osc}}
	     V''(\sigma_{\mathrm{osc}})}{V'(\sigma_{\mathrm{osc}})} - 1 \right)
\right\}^{-1}
 \frac{V'(\sigma_{\mathrm{osc}})}{V'(\sigma_*)}.
\end{equation}

\subsection{Results}
\label{subsec:res}

Putting together the above results, and also adopting 
\begin{equation}
 \rho_{\sigma \mathrm{osc}} = V(\sigma_{\mathrm{osc}}), 
 \label{rhoVsigmaosc}
\end{equation}
we obtain the final
expressions for the density perturbations.
Let us define
\begin{equation}
 X(\sigma_{\mathrm{osc}}) \equiv  
 \frac{1}{2(c-3)} \left(\frac{\sigma_{\mathrm{osc}}
	  V''(\sigma_{\mathrm{osc}})}{V'(\sigma_{\mathrm{osc}})} - 1  \right),
\end{equation}
where $c = 5$ for $t_{\mathrm{reh}} < t_{\mathrm{osc}}$ and 
$c = 9/2$ for $t_{\mathrm{reh}} > t_{\mathrm{osc}}$. 
Note that $X$ denotes effects due to fluctuations of the time when the curvaton
starts its oscillation~$\delta H_{\mathrm{osc}}$.

Then using $\mathcal{P}_{\delta \sigma_*}^{1/2} = H_* / 2 \pi$, 
the linear order density perturbation spectrum is
\begin{equation}
 \mathcal{P}_\zeta = \left(\frac{\partial \mathcal{N}}{\partial \sigma_*} 
  \frac{H_*}{2 \pi} \right)^2
\end{equation}
with 
\begin{equation}
 \frac{\partial \mathcal{N}}{\partial \sigma_*} = \frac{r}{4+3 r}
  \left(1 - X(\sigma_{\mathrm{osc}})\right)^{-1}
 \left\{\frac{V'(\sigma_{\mathrm{osc}})}{V(\sigma_{\mathrm{osc}})} -
  \frac{3 X(\sigma_{\mathrm{osc}})}{\sigma_{\mathrm{osc}}}  \right\}
 \frac{V'(\sigma_{\mathrm{osc}})}{V'(\sigma_*)}.
 \label{NfuncX}
\end{equation}
The second order perturbations in terms of the non-linearity
parameter~$f_{\mathrm{NL}}$ are
\begin{equation}\label{fNLfuncX}
\begin{split}
 f_{\mathrm{NL}} & =
 \frac{5}{6} \frac{\partial^2 \mathcal{N}}{\partial
  \sigma_*^2} 
 \left( \frac{\partial \mathcal{N}}{\partial \sigma_*} \right)^{-2} \\
 & =
  \frac{40 (1+r)}{3 r (4+3 r)}  
 + \frac{5 (4+3 r)}{6 r}
 \left\{\frac{V'(\sigma_{\mathrm{osc}})}{V(\sigma_{\mathrm{osc}})}-
 \frac{3 X(\sigma_{\mathrm{osc}})}{\sigma_{\mathrm{osc}}}  \right\}^{-1}  
  \Biggl[(1-X(\sigma_{\mathrm{osc}}))^{-1} X'(\sigma_{\mathrm{osc}})    \\
 & 
  + \left\{\frac{V'(\sigma_{\mathrm{osc}})}{V(\sigma_{\mathrm{osc}})} -
 \frac{3X(\sigma_{\mathrm{osc}})}{\sigma_{\mathrm{osc}}}  \right\}^{-1} 
 \left\{\frac{V''(\sigma_{\mathrm{osc}})}{V(\sigma_{\mathrm{osc}})} -
 \frac{V'(\sigma_{\mathrm{osc}})^2}{V(\sigma_{\mathrm{osc}})^2} -
 \frac{3 X'(\sigma_{\mathrm{osc}})}{\sigma_{\mathrm{osc}}} + \frac{3
 X(\sigma_{\mathrm{osc}})}{\sigma_{\mathrm{osc}}^2} 
 \right\} \\
 & \qquad \qquad \qquad \qquad \qquad \qquad \qquad \qquad \qquad
  + \frac{V''(\sigma_{\mathrm{osc}})}{V'(\sigma_{\mathrm{osc}})} -
 (1-X(\sigma_{\mathrm{osc}}))
 \frac{V''(\sigma_*)}{V'(\sigma_{\mathrm{osc}})} 
\Biggr].
\end{split}
\end{equation}
Thus we have managed to write down curvature perturbations from
a curvaton in terms of $(r,\, \sigma_{\mathrm{osc}},\, \sigma_*)$,
where $\sigma_{\mathrm{osc}}$ is determined by $\sigma_*$ through
(\ref{sigmaosc}). 

The curvaton energy fraction upon decay~$r$ (\ref{efr}) can
be estimated by assuming that the universe suddenly switches between
matter/radiation domination at reheating (which happens when $H=
\Gamma_\phi$, with $\Gamma_\phi$ the inflaton decay rate) 
and also when the curvaton dominates the universe (if it ever does), 
\begin{equation}\label{anar}
\begin{split}
 r & = \mathrm{Max.} \Biggl[
\frac{\rho_{\sigma\mathrm{osc}}}{\rho_{\phi \mathrm{reh}}}
  \frac{H_{\mathrm{reh}}^2 }{H_{\mathrm{osc}}^{3/2} H_{\mathrm{dec}}^{1/2} }
  \times  \mathrm{Min.} \left(
1, \,  \frac{H_{\mathrm{reh}}^{1/2}}{H_{\mathrm{osc}}^{1/2}}
\right),
 \\
& \qquad  \qquad \left\{
\frac{\rho_{\sigma\mathrm{osc}}}{\rho_{\phi \mathrm{reh}}}
  \frac{H_{\mathrm{reh}}^2 }{H_{\mathrm{osc}}^{3/2} H_{\mathrm{dec}}^{1/2} }
  \times  \mathrm{Min.} 
\left(
1, \,  \frac{H_{\mathrm{reh}}^{1/2}}{H_{\mathrm{osc}}^{1/2}}
\right)
\right\}^{4/3}
\Biggr]
 \\
& = \mathrm{Max.}\Biggl[
\frac{V(\sigma_{\mathrm{osc}})}{3 M_p^2 H_{\mathrm{osc}}^{3/2}
 \Gamma_\sigma^{1/2}} \times \mathrm{Min.}\left(
1, \,  \frac{\Gamma_\phi^{1/2}}{H_{\mathrm{osc}}^{1/2}} 
\right),
\\
& \qquad \qquad 
\left\{
\frac{V(\sigma_{\mathrm{osc}})}{3 M_p^2 H_{\mathrm{osc}}^{3/2}
 \Gamma_\sigma^{1/2}} \times \mathrm{Min.}
\left(
1, \,  \frac{\Gamma_\phi^{1/2}}{H_{\mathrm{osc}}^{1/2}} 
\right) 
\right\}^{4/3}
\Biggr],
\end{split}
\end{equation}
Note that 
we have again especially made use of Assumption~\ref{ass4}.
The first and second terms in the $\mathrm{Max.}$ parentheses
correspond to the curvaton being subdominant and dominant at decay,
respectively. Also, the $\mathrm{Min.}$ parentheses are due to whether
the onset of oscillation is after or before reheating.

\subsubsection*{example: Quadratic Potential}

As a simple example, let us consider the familiar case of a quadratic
potential $V \propto \sigma^2$. One sees that $X$ vanishes,
i.e. the onset of the curvaton oscillation is uniform, and 
\begin{equation}
 \frac{\partial \mathcal{N}}{\partial \sigma_*} = \frac{2r}{4+3r} \frac{1}{\sigma_*},
 \qquad 
 f_{\mathrm{NL}} = \frac{5}{12} \left(
 -3 + \frac{4}{r} + \frac{8}{4+3 r}\right),
 \label{quadcurvaton}
\end{equation}
where the well studied results are reproduced. 
Here, in particular, large non-Gaussianity $f_{\mathrm{NL}} \gg 1 $ is
sourced only by curvatons decaying while $ r \ll 1$.

\subsection{Validity of the Analytic Expressions}
\label{sec:val}

Upon deriving the above analytic results, we have adopted simplifying assumptions
such as sudden decays, and also that the
curvaton suddenly starts a sinusoidal oscillation when
(\ref{dotsigma}). 
Here we comment on the validity of such assumptions and the analytic results.

First we should remark that upon obtaining the expression~(\ref{2424})
for~$H_{\mathrm{osc}}$, we have assumed the curvaton to follow the 
attractor solution~(\ref{eq17}) until the onset of the
oscillation~(\ref{dotsigma}). As is discussed in the appendix, this
procedure is validated under the necessary condition~(\ref{sscond}). 
Thus one should be careful when dealing with curvaton potentials along which
(\ref{sscond}) breaks down significantly before the oscillation.
There may also be cases where (\ref{sscond}) once breaks down but
recovers afterwards, such that the curvaton once deviates from the 
approximation~(\ref{eq17}) but comes back again later on, 
if, say, the curvaton potential possesses a plateau region. 
However, we also note that the breakdown of~(\ref{sscond}) before
the onset of the oscillation does not necessary imply the breakdown of
the analytic results derived above. 
When $|V''/H^2|$ becomes of order unity, the parameter~$c$ in the
approximation~(\ref{eq17}) no longer takes constant
values~(\ref{ciroiro}), but becomes time-dependent 
as can be seen from~(\ref{cVpp}). 
Hence using the time-dependent solution of (\ref{cVpp})
for~$c$ (given that the solution exists\footnote{Note that (\ref{cVpp}) admits
solutions of~$c$ for largely negative~$V''/H^2$, but has no solutions
for largely positive~$V''/H^2$.})
may give more accurate estimations, 
but as long as the time scale of the variation of~$c$ is of order the
Hubble time, our discussions with constant~$c$ 
should not be drastically modified. 
Moreover, unless (\ref{sscond}) breaks down long before the onset
of the oscillations, the deviation of the curvaton dynamics from
(\ref{eq17}) with constant~$c$ is mild and the above analytic results
stay valid.
We will see in Subsection~\ref{subsubsec:NGht} an explicit example where
the condition~(\ref{sscond}) clearly breaks down by the time when the
oscillation starts, but still the above analytic results giving good
estimations of the density perturbations.

Upon estimating $\rho_{\sigma \mathrm{osc}}$, we have
adopted~(\ref{rhoVsigmaosc}) and ignored the curvaton's kinetic
energy. This procedure is validated in many cases where the potential
energy is larger than the kinetic energy at the
onset of the oscillations. Moreover, as long as the ratio between the
potential and kinetic energies is a constant, considerations of the
kinetic energy drop out through $\delta \ln \rho_{\sigma
\mathrm{osc}}$.
However, there may be cases for which it becomes important to take into
account the curvaton's kinetic energy at~$t_{\mathrm{osc}}$. 

It was also assumed that even if the curvaton potential 
deviates from a quadratic one (giving rise to~$\delta
H_{\mathrm{osc}}$),
once the oscillation starts its amplitude soon decays and the curvaton
starts a sinusoidal oscillation about its minimum where the potential is
well approximated by a quadratic one. 
However, we should also note that if the curvaton oscillation starts
with an amplitude large enough so that its potential deviates
from quadratic for a while after~$t_{\mathrm{osc}}$, then
the curvaton energy density does not behave as matter but
decays as $\rho_\sigma \propto a^{-n}$ with $n \neq 3$.  
This can modify our analytic expressions, in addition to changing~$r$ and
thus the amplitude of the perturbation spectrum.
In cases where the oscillation history is rather trivial (e.g. $V =
\frac{1}{2} m^2 \sigma^2 + \lambda \sigma^4$ giving $n = 4$ $ \to$
$n=3$), extending the discussions above is rather straightforward, as is
shown in Appendix~\ref{app:non-sinu}.
However for more complicated cases (e.g. curvaton oscillation in a
potential with superimposed periodic modulations, which can give rise to a
time-dependent~$n$), some parts of our results may require major corrections.
We leave this for future work.

\subsection{Spectral Index}
\label{sec:SI}

Although the perturbation amplitude crucially depends on the curvaton
dynamics after inflation, the spectral index
\begin{equation}
 n_s - 1 \equiv \frac{d \ln \mathcal{P}_\zeta}{d \ln k}
\end{equation}
is determined only by the information at the horizon exit during inflation. 
Note that the linear order density perturbation is expressed by
\begin{equation}
 \frac{\partial \mathcal{N}}{\partial \sigma_*} = 
 \left( \mathrm{function\, \, of \, \, } r,\,
  \sigma_{\mathrm{osc}}\right)\times
 \frac{1}{V'(\sigma_*)},
\end{equation}
where $r$ and $\sigma_{\mathrm{osc}}$ are independent of the comoving
wave number~$k$. Hence
\begin{equation}
 n_s - 1 = \frac{d}{d\ln k} \ln \left(\frac{H_*}{V'(\sigma_*)}\right)^2
  \simeq
 \frac{2}{3} \frac{V''(\sigma_*)}{H_*^2} + 2 \frac{\dot{H}_*}{H_*^2},
 \label{ns-1}
\end{equation}
where we have made use of $d\ln k \simeq H_* dt$ and $3 H_*
\dot{\sigma}_* \simeq -V'(\sigma_*)$. One clearly sees that for
inflation with $H_* = \mathrm{const.}$, the observationally suggested
red-tilted perturbation spectrum requires the curvaton to be tachyonic
during inflation. See also discussions in Footnote~\ref{foot1}.

\section{Curvatons Along Flat Potentials}
\label{sec:flat}

The results (\ref{NfuncX}) and (\ref{fNLfuncX}) in the previous section 
show that additional contributions to the density perturbations
from~$\delta H_{\mathrm{osc}}$ can suppress/enhance the
perturbation amplitude as well as the non-Gaussianity. 
As an interesting case where the $\delta H_{\mathrm{osc}}$~contributions
to the density perturbations become important, in this section 
we focus on a curvaton starting its oscillation from a flat region 
where the potential tilt is tiny.

Given that the first derivative of the potential at the onset of
oscillation is small to enough to satisfy
\begin{equation}
 \left|\frac{\sigma_{\mathrm{osc}}
  V''(\sigma_{\mathrm{osc}})}{V'(\sigma_{\mathrm{osc}})}\right|, 
 \,
 \left|\frac{V''(\sigma_{\mathrm{osc}})
  V(\sigma_{\mathrm{osc}})}{V'(\sigma_{\mathrm{osc}})^2}  \right|, 
 \,
 \left|\frac{V''(\sigma_{\mathrm{osc}})^2}{V'(\sigma_{\mathrm{osc}})
  V'''(\sigma_{\mathrm{osc}})}  \right|, 
 \,
 \left|\frac{\sigma_{\mathrm{osc}}
  V''(\sigma_*)}{V'(\sigma_{\mathrm{osc}})}  \right| 
 \gg 1,
 \label{gg}
\end{equation}
then one finds
\begin{equation}
 |X(\sigma_{\mathrm{osc}})| \simeq \left|\frac{1}{2 (c-3)}
	     \frac{\sigma_{\mathrm{osc}}
	     V''(\sigma_{\mathrm{osc}})}{V'(\sigma_{\mathrm{osc}})}
	    \right|   \gg 1, \quad
 X'(\sigma_{\mathrm{osc}}) \simeq -\frac{1}{2
 (c-3)}\frac{\sigma_{\mathrm{osc}}V''(\sigma_{\mathrm{osc}})^2}{
 V'(\sigma_{\mathrm{osc}})^2},   
\end{equation}
and the equations (\ref{NfuncX}) and (\ref{fNLfuncX}) are approximated by
\begin{equation}
 \frac{\partial \mathcal{N}}{\partial \sigma_*} \simeq \frac{3 r}{4+3 r}
  \frac{V'(\sigma_{\mathrm{osc}})}{\sigma_{\mathrm{osc}} V'(\sigma_*)}, 
 \label{Nflat}
\end{equation}
\begin{equation}
 f_{\mathrm{NL}} \simeq -\frac{5 (4+ 3r )}{18 r}
  \frac{\sigma_{\mathrm{osc}}
  V''(\sigma_*)}{V'(\sigma_{\mathrm{osc}})}. 
 \label{fNLflat}
\end{equation}
Note that here, $|f_{\mathrm{NL}}| \gg 1$ is guaranteed (even when $r
\gg 1$). Furthermore, when
$V(\sigma)$ is a monotonically increasing function around its origin,
i.e. $V' > (<) 0$ for 
$\sigma > (<) 0$, then $f_{\mathrm{NL}} \propto  (\mathrm{negative}\,
\, \mathrm{factor}) \times V''(\sigma_*)$. In other words, 
with constant~$H$ inflation, 
a red(blue)-tilted perturbation spectrum is always accompanied
by a largely positive (negative) $f_{\mathrm{NL}}$. 

We should remark that under the condition $|\sigma V'' / V' | \gg 1$ in (\ref{gg}),
(\ref{sscond}) no longer holds at~$t_{\mathrm{osc}}$.
This may lead to some deviation of the curvaton dynamics from the
approximation~(\ref{eq17}), giving rise to errors in the estimations. 
However, as we will see in Section~\ref{sec:PNGC} when we study explicit
cases, the overall behaviour of the resulting density perturbations
discussed here are still valid as long as the breakdown of the
approximation~(\ref{eq17}) is mild. (See also discussions in
Subsection~\ref{sec:val}).

\subsection{Hilltop Curvatons}
\label{subsec:hilltop}

Let us further focus on a curvaton located at the hilltop, i.e. a curvaton whose
potential is well approximated by
\begin{equation}
 V(\sigma) = V_0 - \frac{1}{2} m^2 (\sigma - \sigma_0)^2
\end{equation}
around $\sigma_{\mathrm{osc}}$ and $\sigma_*$, where $m$,
$\sigma_0$, and $V_0 (>0)$ are constants. Such case will be important
when we study pseudo-Nambu-Goldstone curvatons in the next section. 
Without loss of generality, hereafter we assume $0 <
\sigma_{\mathrm{osc}} < \sigma_* < \sigma_0$. 

Now the flatness conditions in (\ref{gg}) translate into
\begin{equation}
 \frac{\sigma_{\mathrm{osc}}}{\sigma_0 - \sigma_{\mathrm{osc}}} \gg 1,
 \quad
 \frac{V_0}{m^2 (\sigma_{\mathrm{osc}} - \sigma_0)^2 } \gg 1,
 \label{eq36}
\end{equation}
under which (\ref{Nflat}) and (\ref{fNLflat}) are
\begin{equation}
 \mathcal{P}_\zeta^{1/2} \simeq \frac{3r}{4+3 r}\frac{\sigma_0 -
  \sigma_{\mathrm{osc}}}{\sigma_0 - \sigma_*} \frac{H_{\mathrm{inf}}}{2
  \pi \sigma_{\mathrm{osc}}},  \label{P}
\end{equation}
\begin{equation}
 f_{\mathrm{NL}} \simeq \frac{5 (4+3 r)}{18 r}
  \frac{\sigma_{\mathrm{osc}}}{\sigma_0 - \sigma_{\mathrm{osc}}},  \label{fNL}
\end{equation}
with spectral index
\begin{equation}
 n_s -1 = -\frac{2}{3} \frac{m^2}{H_{\mathrm{inf}}^2} < 0.
\end{equation}
Here we note that throughout this subsection, inflation with
constant~$H$ is assumed. 

\begin{figure}[htbp]
\begin{center}
  \begin{center}
 \includegraphics[width=.55\linewidth]{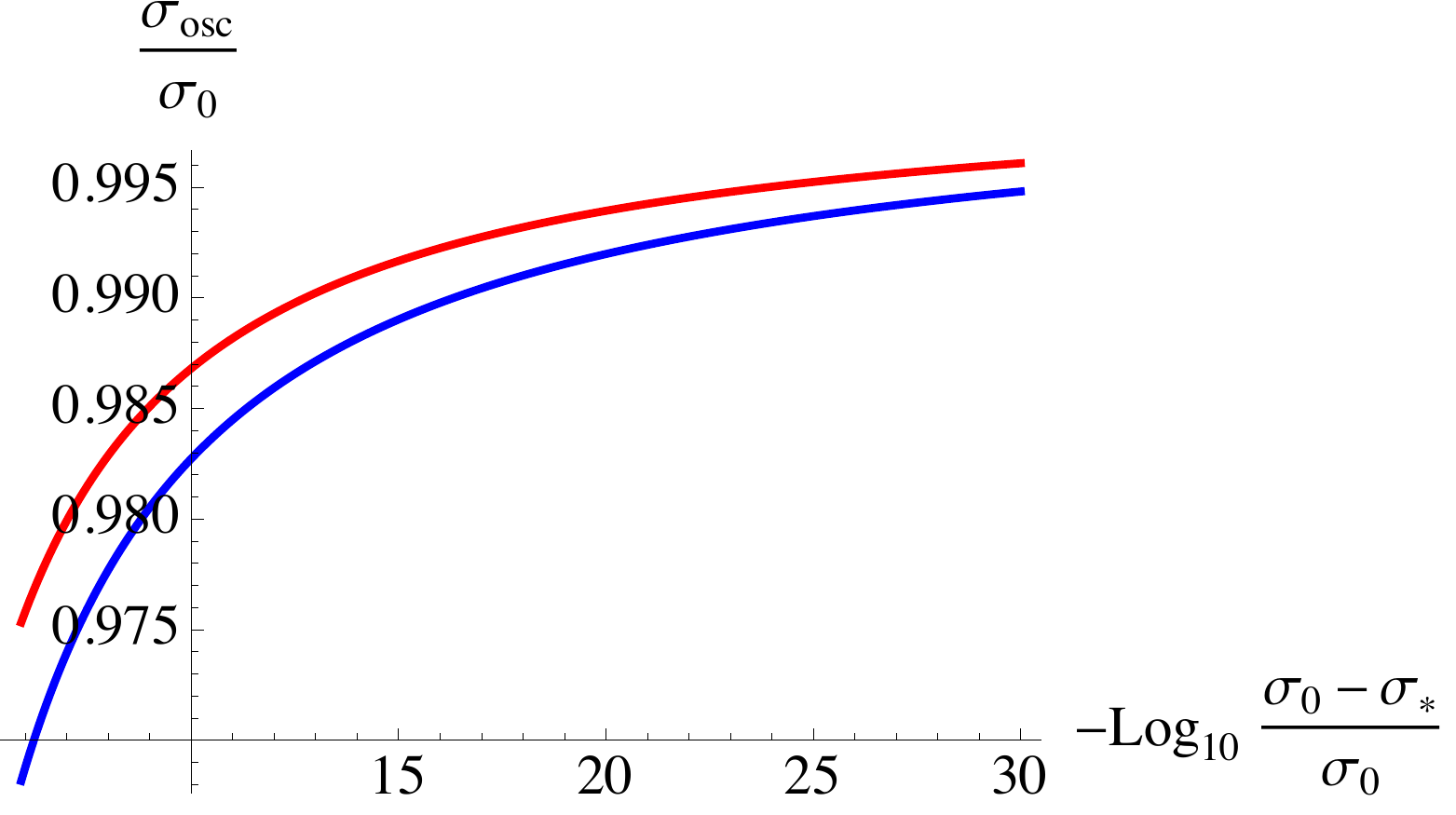}
  \end{center}
  \caption{$\sigma_{\mathrm{osc}}$ in terms of $\sigma_*$, obtained from
 (\ref{eq311}) with $c = 5$ (red) and $c =  9/2$ (blue).} 
  \label{fig:sigma_osc}
\end{center}
\end{figure}

The Hubble parameter at the onset of oscillation~(\ref{2424}) is 
\begin{equation}
 H_{\mathrm{osc}}^2 = \frac{m^2 (\sigma_0 - \sigma_{\mathrm{osc}}) }{c
  \sigma_{\mathrm{osc}}} \ll H_{\mathrm{inf}}^2,
\end{equation}
where the inequality follows from (\ref{eq36}) and $|n_s -1| = \mathcal{O}(10^{-2})$.
Thus $\sigma_*$ and $\sigma_{\mathrm{osc}}$ can be related by
solving~(\ref{sigmaosc}), which in the 
hilltop limit, i.e. $\sigma_* \to \sigma_0$ (from below), gives
\begin{equation}
 \ln \left( \frac{\sigma_0 - \sigma_*}{\sigma_0 - \sigma_{\mathrm{osc}}} \right)
 \simeq 
 - \frac{1}{2 (c-3) } \frac{\sigma_{\mathrm{osc}}}{\sigma_0
       - \sigma_{\mathrm{osc}}} , 
 \label{eq311}
\end{equation}
where $c = 5$ for $t_{\mathrm{reh}} < t_{\mathrm{osc}}$ and 
$c = 9/2$ for $t_{\mathrm{reh}} > t_{\mathrm{osc}}$.
Since the left hand side is logarithmic, one sees that
$\sigma_{\mathrm{osc}}$ approaches $\sigma_0$ much slower than
$\sigma_*$ does, cf. Figure~\ref{fig:sigma_osc}.
Thus, in the hilltop limit, $\mathcal{P}_\zeta$ (\ref{P}) blows
up due to the enhancement factor $(\sigma_0 - \sigma_{\mathrm{osc}}) /
(\sigma_0 - \sigma_*)$, while $f_{\mathrm{NL}}$ (\ref{fNL}) increases
slowly. 
One can see from this example that the evaluation of
$\sigma_{\mathrm{osc}}$ is crucial for correctly estimating the density
perturbations. 
The extreme amplification of the perturbation amplitude can be
understood as curvatons taking longer time to start its oscillation 
as one approaches the hilltop limit. 

\vspace{\baselineskip}

One may worry, in the hilltop limit, that the curvaton during inflation
jumps over the hill due to quantum fluctuations $\Delta \sigma  = \pm
H_{\mathrm{inf}}/ 2 \pi$, and domain walls be generated.
Here we note that if the hilltop curvaton generates density
perturbations that satisfy the 
observational constraints from the WMAP 7-year
data~\cite{Larson:2010gs} $\mathcal{P}_\zeta \approx 2.4 \times 10^{-9}$ and 
$f_{\mathrm{NL}} < 74$ (for local-type non-Gaussianity which is
important for us here), the condition for the absence of domain walls is
satisfied, 
\begin{equation}
 \frac{H_{\mathrm{inf}}}{2 \pi} (\sigma_0 - \sigma_*)^{-1}
 \simeq \frac{6}{5} \mathcal{P}_\zeta^{1/2}
  f_{\mathrm{NL}} \lesssim 0.004.
\end{equation}
We also have to check whether quantum fluctuations lead to a
random walk of the curvaton.
Further using the WMAP7 central value $n_s \approx 0.96$, one sees that
the classical rolling dominates over the quantum fluctuations,
\begin{equation}
 \frac{H_{\mathrm{inf}}}{2 \pi}
 \left| \frac{\dot{\sigma_*}}{H_{\mathrm{inf}}} \right| ^{-1} =
 \frac{3}{2 \pi} \frac{H_{\mathrm{inf}}^3}{V'(\sigma_*)} \simeq
  \frac{12}{5} \frac{\mathcal{P}_\zeta^{1/2} f_{\mathrm{NL}}}{1-n_s} 
 \lesssim 0.2.
\end{equation}

Since the onset of the curvaton oscillation is delayed as one approaches
the hilltop limit, it becomes easier for the curvaton to dominate the
universe before it decays. However, we should also 
note that the curvaton energy fraction when it starts to oscillate,
\begin{equation}
 \frac{\rho_{\sigma \mathrm{osc}}}{3 M_p^2 H_{\mathrm{osc}}^2} \simeq
 \frac{3r}{4+3 r} \frac{4 c}{5} \frac{f_{\mathrm{NL}}}{1 - n_s}
 \frac{V_0}{3 M_p^2 H_{\mathrm{inf}}^2} 
 \lesssim
  \frac{3r}{4+3 r} 
 \frac{V_0}{3 M_p^2 H_{\mathrm{inf}}^2} 
 \times 7000 ,
 \label{7000}
 \end{equation}
can become as large as unity.
In such case, the above estimations can break down
since upon deriving the analytical expressions, we have ignored the
effects of the curvaton energy density on the evolution of the universe
before the curvaton starts its oscillation, cf. Assumption~\ref{ass4}.

\section{Case Study: Pseudo-Nambu-Goldstone Curvatons}
\label{sec:PNGC}

As a simple example of a curvaton model, let us examine the case where
the curvaton is a pseudo-Nambu-Goldstone (NG) boson of a broken U(1)
symmetry. The approximate symmetry suppresses the curvaton mass, while
the periodicity of the U(1) provides both (local) minima and
maxima along the curvaton potential, thus allowing the curvaton to
produce a blue-tilted as well as a red-tilted density perturbation
spectrum. 

In particular, we study the NG curvaton potential of the form
\begin{equation}
 V(\sigma) = \Lambda^4 \left[ 1 - \cos \left(\frac{\sigma}{f}\right)
		       \right],
 \label{eq:NGpot}
\end{equation}
where $f$ and $\Lambda$ are mass scales.
Without loss of generality, we set the field value of the curvaton at
horizon exit during inflation to lie within the range $ 0 < \sigma_* <
\pi f$, and consider its oscillation about the origin~\mbox{$\sigma = 0$}. 
Supposing that the coupling of the NG curvaton with its decay product is
suppressed by the scale of symmetry breaking~$f$, then the decay rate of the
curvaton takes the value
\begin{equation}
 \Gamma_\sigma = \frac{\beta}{16 \pi} \frac{m^3}{f^2}
 =  \frac{\beta}{16 \pi} \frac{\Lambda^6}{f^5}, \label{beta}
\end{equation}
where $m$ is the mass at the minimum, i.e. $m^2 = V''(0)$, and $\beta$
is a constant whose precise value depends on the decay modes.
Throughout this section we fix $\beta$ to unity. 
(Cases with smaller~$\beta$ will be discussed towards the end of
Subsection~\ref{subsub:windows}). 
Furthermore, we consider inflation with constant~$H$, and
assume the pivot scale to have exited the horizon 50 e-foldings
before the end of inflation. 

We will see that an NG curvaton away from the hilltop
requires inflation and the reheating scales to lie within a narrow range
in order to generate density perturbations with spectral index of order
$n_s - 1 \sim -10^{-2}$. 
However, we also show that as one approaches the hilltop limit $\sigma_*
\to \pi f$, the allowed inflation/reheating scales extremely broaden 
and that the non-Gaussianity lies in the range 
$10 \lesssim f_{\mathrm{NL}} \lesssim 30$.

For an investigation of NG curvatons located close to their potential
minimum, see also~\cite{Dimopoulos:2003az}.

\subsection{Density Perturbations from NG Curvatons}

Let us first investigate density perturbations generated by
NG curvatons, and compare the analytic estimations in the previous
sections with numerically computed results. 
We start from NG curvatons located away from the hilltop of the
potential, then move on to the hilltop limit.

\subsubsection{Non-Hilltop Region}

Firstly, we choose the potential parameters and the inflation/reheating scales
such that the curvaton with $\sigma_* = \frac{3}{4} \pi f$ 
generates a density perturbation spectrum whose amplitude takes the
COBE normalization value
\begin{equation}
 \mathcal{P}_\zeta \approx 2.4 \times 10^{-9},   \label{COBE}
\end{equation}
with the spectral index lying at the central value of the
WMAP7~\cite{Larson:2010gs}  bound 
(for a power-law spectrum with no tensor modes)
\begin{equation}
 n_s \approx 0.96. \label{7ns}
\end{equation}
Then, considering the case where $t_{\mathrm{osc}} < t_{\mathrm{reh}}$,
and that the curvaton dominates the universe before it decays, 
we adopt the following parameter set (which lies at around the center of
the allowed window in Figure~\ref{fig:75dom} in the next subsection):
$H_{\mathrm{inf}} = 10^{13.5}\, \mathrm{GeV}$, 
$\rho_{\mathrm{reh}}^{1/4} = 10^{15}\, \mathrm{GeV}$
(i.e. $\Gamma_\phi \approx 9.73\times 10^{-8} M_p$),
$f \approx 3.36 \times 10^{-2} M_p$, and $\Lambda \approx 3.56 \times 10^{-4} M_p$.

\vspace{\baselineskip}

\begin{figure}[htbp]
\begin{center}
  \begin{center}
 \includegraphics[width=.5\linewidth]{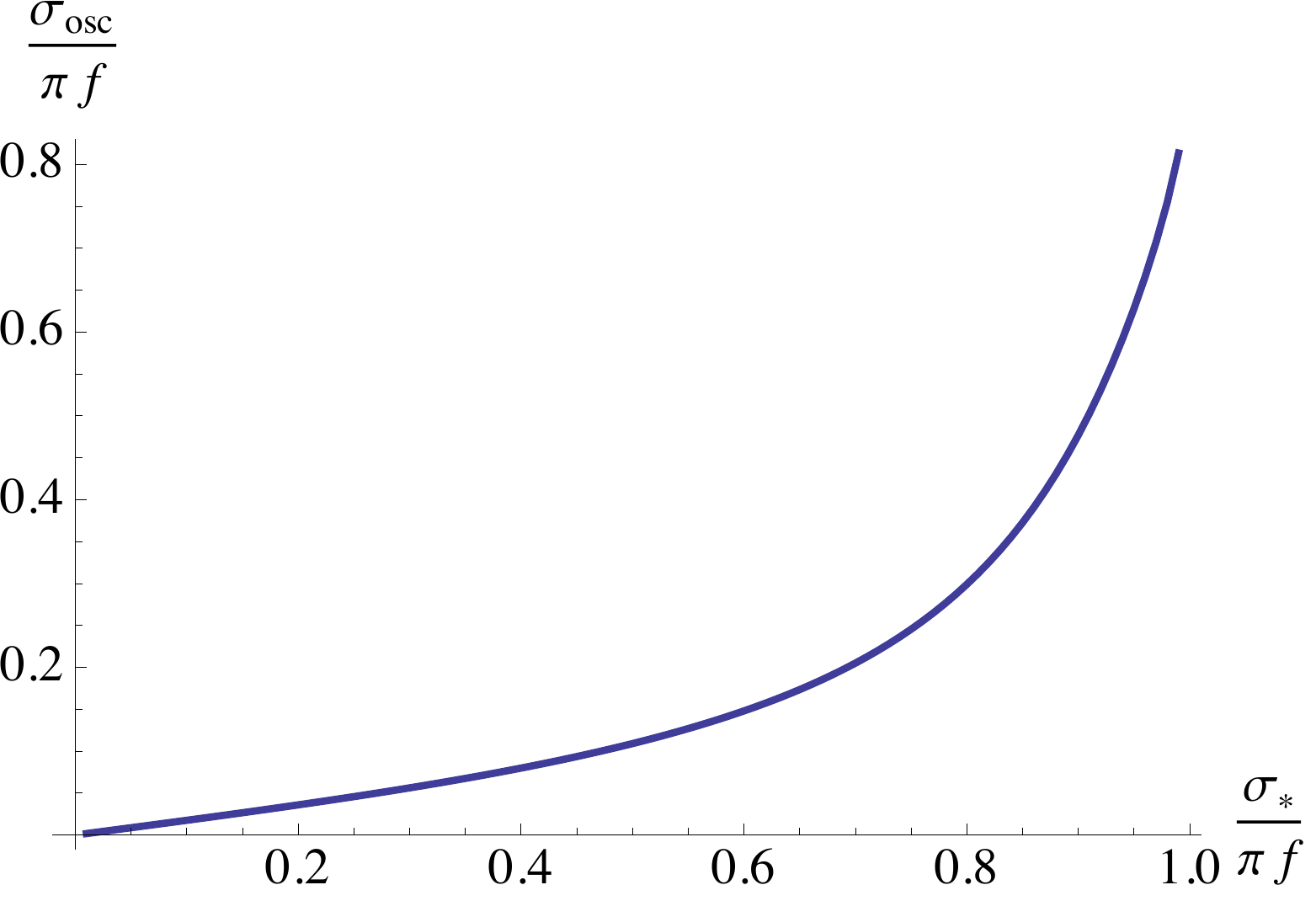}
  \end{center}
  \caption{$\sigma_{\mathrm{osc}}$ in terms of $\sigma_*$ for the
 NG curvaton, cf.~(\ref{NGfit}).} 
  \label{fig:NG-sigma_osc}
\end{center}
\end{figure}

In order to analytically estimate the density perturbations, we need to solve 
(\ref{sigmaosc}) for obtaining $\sigma_{\mathrm{osc}}$.
Introducing
\begin{equation}
 \alpha \equiv \frac{\sigma}{\pi f},
\end{equation}
one obtains
\begin{equation}
 \ln \left|
 \frac{\tan \left(\alpha_{\mathrm{osc}} \pi/2\right)}{\tan
 \left(\alpha_* \pi/2\right)}
  \right| = 
 - \frac{\mathcal{N}_*}{3 H_{\mathrm{inf}}^2} \frac{\Lambda^4}{f^2} - 
 \frac{1}{2 (c-3)} \frac{\alpha_{\mathrm{osc}} \pi }{\sin
 (\alpha_{\mathrm{osc}} \pi )},
 \label{4242}
\end{equation}
whose solution for $\mathcal{N}_* = 50$ and $c = 9/2$ in the region
$ 0.01 \lesssim \alpha_* \lesssim 0.99$
is approximated by (cf. Figure~\ref{fig:NG-sigma_osc})
\begin{equation}
\alpha_{\mathrm{osc}} \approx 
 0.195 \alpha_* - 0.0993 \alpha_*^2 +0.846 \alpha_*^4 - 1.34 \alpha_*^6
 + 1.26 \alpha_*^8.
 \label{NGfit}
\end{equation}
Then by plugging (\ref{anar}), (\ref{eq:NGpot}), and (\ref{NGfit}) into
(\ref{NfuncX}) and (\ref{fNLfuncX}), one obtains the linear order density
perturbations and its non-Gaussianity as functions of $\alpha_* =
\sigma_* / \pi f$. 

\vspace{\baselineskip}

We have also numerically computed the density perturbations by solving
the equation of motion of the curvaton. 
After 50~e-foldings of inflation with constant~$H$ since the exit of the
pivot scale, the inflaton energy is transferred to inflaton matter decaying
as $\rho_m \propto a^{-3}$, which eventually decays into radiation
($\rho_r \propto a^{-4}$).
Sudden decays are adopted for the inflaton and the curvaton, 
i.e. they instantly decay into radiation when $H = \Gamma$.

We run the numerical computations with different $\sigma_*$
(the curvaton velocity at the horizon exit is set to the slow-roll
attractor value, which is uniquely determined by~$\sigma_*$), and
evaluate the number of e-foldings obtained by the time the curvaton
decays. Thus evaluating the 
differences in e-folding numbers~$\Delta \mathcal{N}$ due to shifts in
the initial condition~$\Delta \sigma$, we numerically obtain $\partial
\mathcal{N} / \partial \sigma_*$. Second order perturbations as well as the
spectral index are computed in similar manners.

\vspace{\baselineskip}

\begin{figure}[htbp]
 \begin{minipage}{.48\linewidth}
  \begin{center}
 \includegraphics[width=\linewidth]{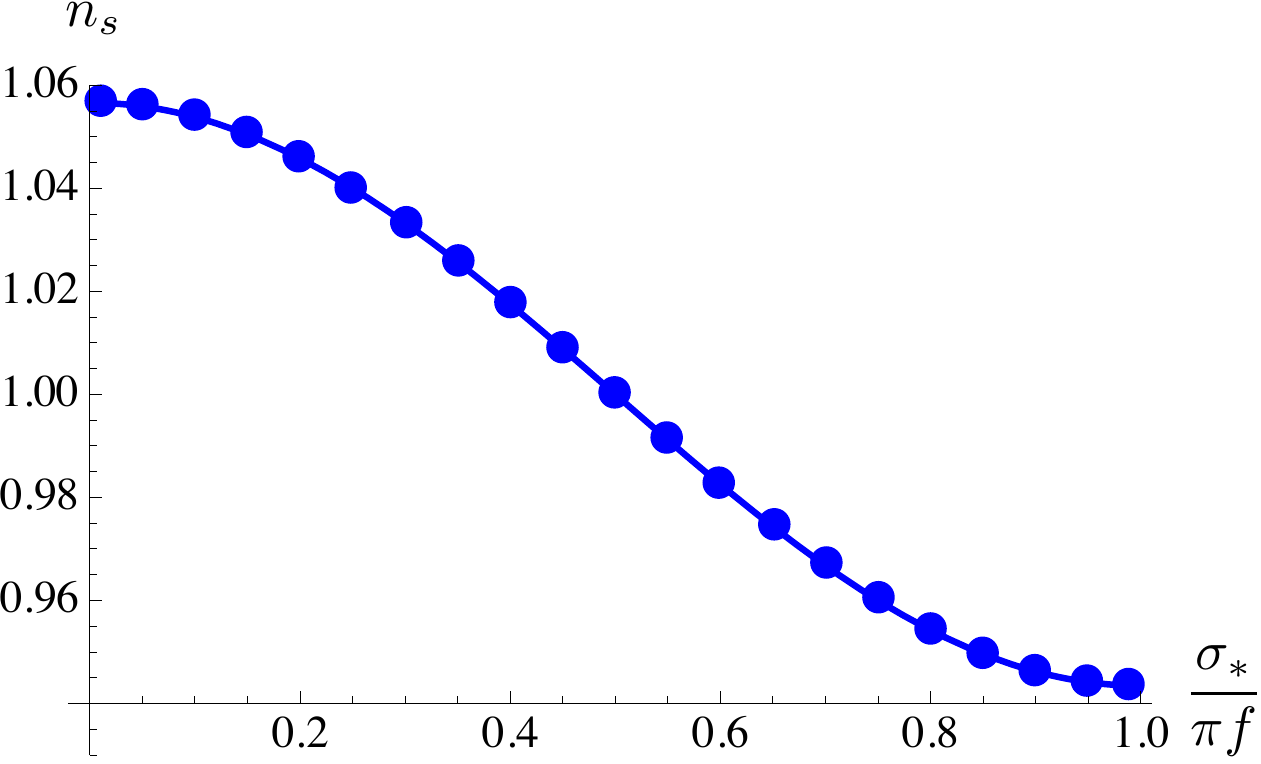}
  \end{center}
  \caption{Spectral index.}
  \label{fig:NG-ns}
 \end{minipage} 
 \begin{minipage}{0.01\linewidth} 
  \begin{center}
  \end{center}
 \end{minipage} 
 \begin{minipage}{.48\linewidth}
  \begin{center}
 \includegraphics[width=\linewidth]{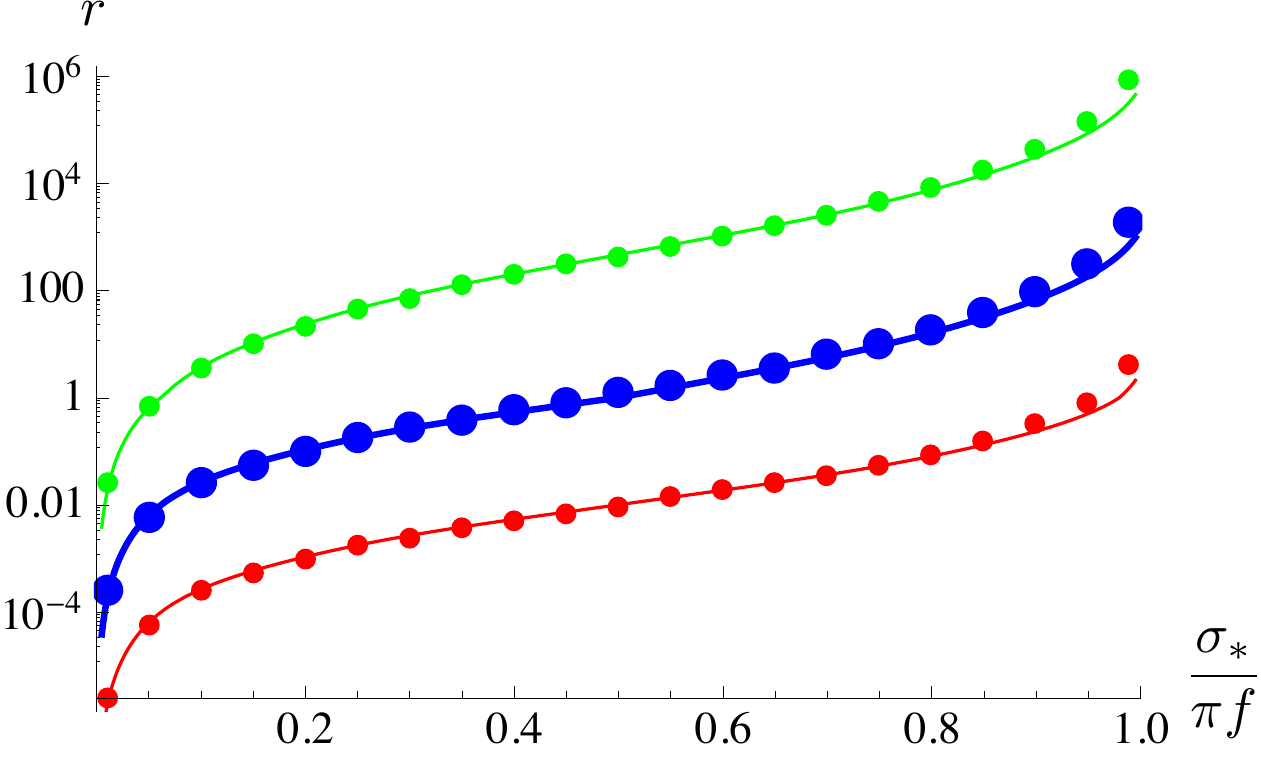}
  \end{center}
  \caption{Energy fraction at decay.}
  \label{fig:NG-r}
 \end{minipage} 
\end{figure}

\begin{figure}[htbp]
 \begin{minipage}{.48\linewidth}
  \begin{center}
  \includegraphics[width=\linewidth]{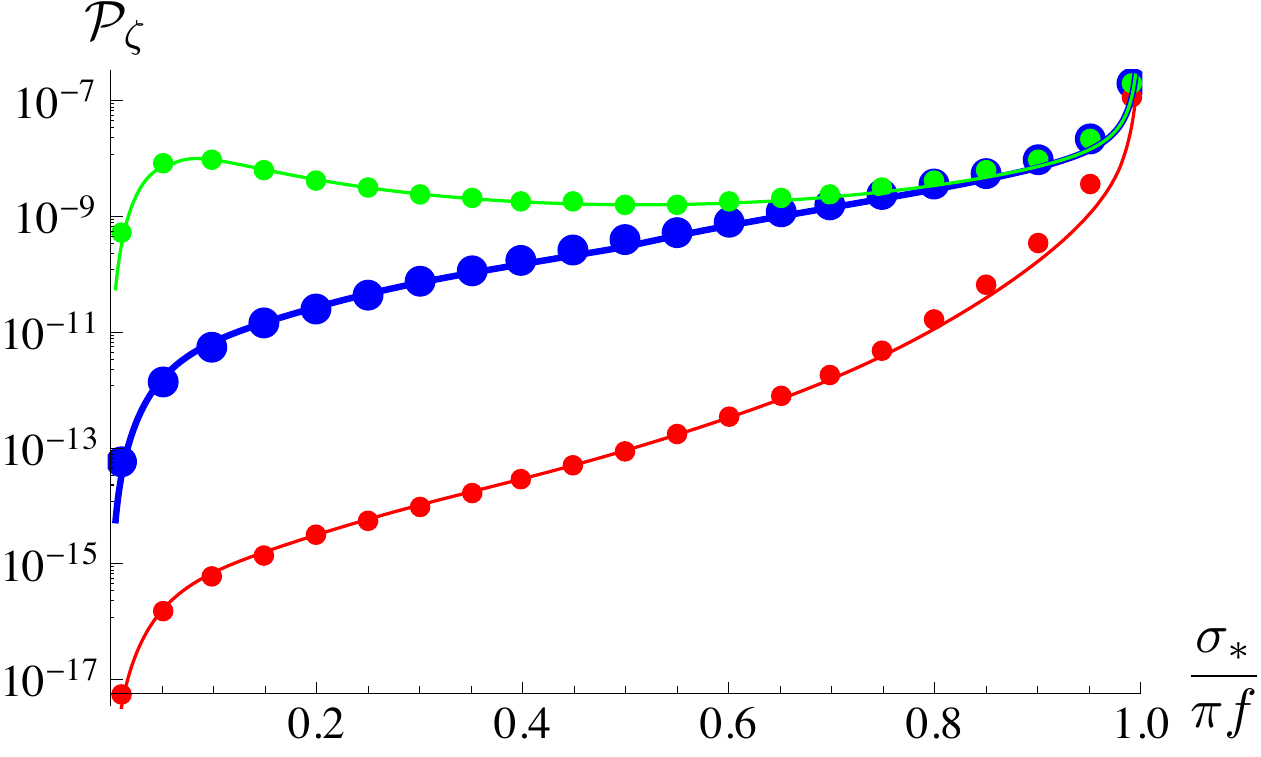}
  \end{center}
  \caption{Power spectrum.}
  \label{fig:NG-P}
 \end{minipage} 
 \begin{minipage}{0.01\linewidth} 
  \begin{center}
  \end{center}
 \end{minipage} 
 \begin{minipage}{.48\linewidth}
  \begin{center}
 \includegraphics[width=\linewidth]{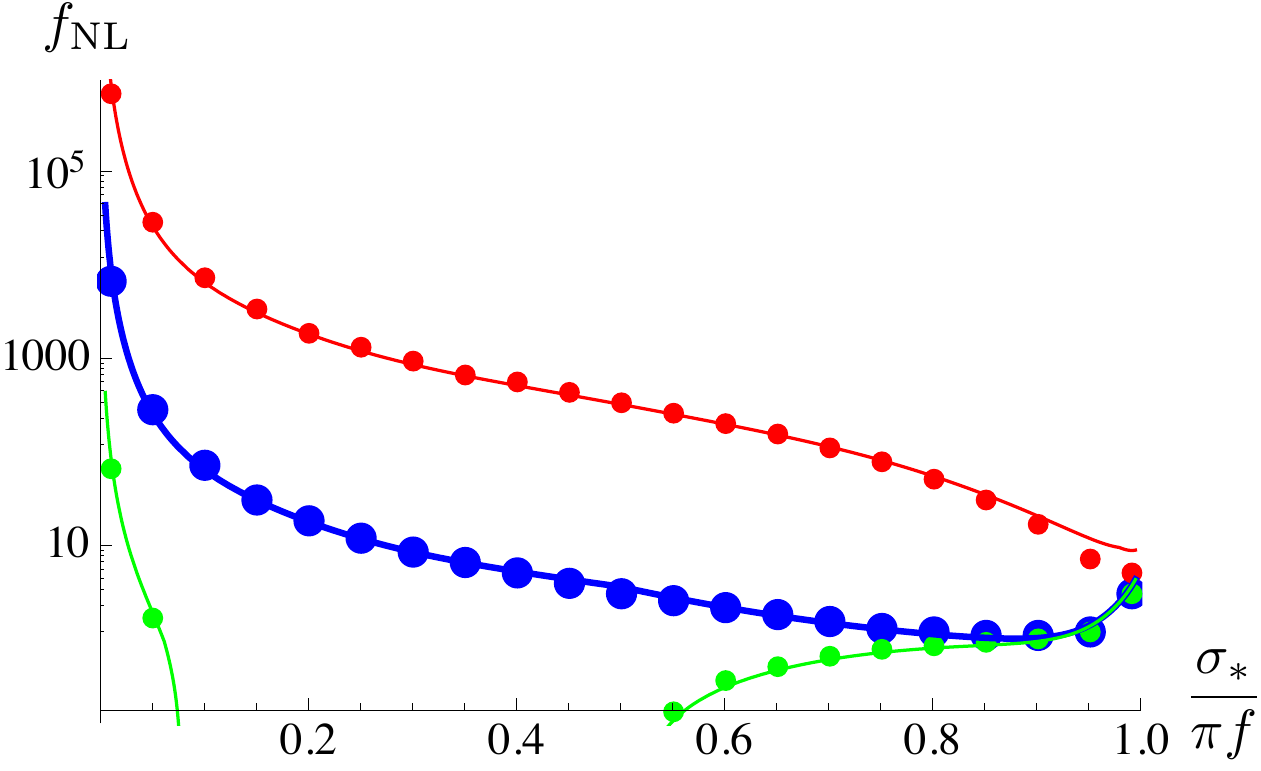}
  \end{center}
  \caption{Non-Gaussianity. (The green line goes negative in the
  covered region.)}
  \label{fig:NG-fNL}
 \end{minipage} 
\end{figure}

In Figures~\ref{fig:NG-ns} - \ref{fig:NG-fNL} we lay out the results in
terms of $\sigma_*$, where the solid lines are the analytic estimations
and the dots are the numerical results. 
Blue lines and dots correspond to our case with $\Gamma_\sigma = \frac{1}{16 \pi}
\frac{m^3}{f^2} $ where 
the WMAP values (\ref{COBE}) and (\ref{7ns}) are realized at $\alpha_* =
3/4$, 
and for comparison we also show cases with 
$\Gamma_\sigma = 10^{4} \times \frac{1}{16 \pi}\frac{m^3}{f^2}$ (red) and 
$\Gamma_\sigma = 10^{-4} \times \frac{1}{16 \pi}\frac{m^3}{f^2}$
(green). (Note that Figure~\ref{fig:NG-ns} has only one line since the
spectral index is independent of the decay rate.) 
One sees that the results asymptote to that of the familiar quadratic
curvaton~(\ref{quadcurvaton}) as $\alpha_* \to 0$.
On the other hand, the linear order perturbation as well as the
non-Gaussianity increases as one approaches $\alpha_* \to 1$, 
smoothly connecting to the hilltop limit which we will soon study. 
The curvaton energy fraction at decay~$r$ increases with~$\alpha_*$, and
the density perturbations become $r$-independent for $r \gg 1$, as can
be seen from the overlap of the blue and green lines towards $\alpha_* \to 1$.


\subsubsection{Hilltop Region}
\label{subsubsec:NGht}

As we have seen in Subsection~\ref{subsec:hilltop}, an extreme enhancement
of the density perturbations occurs in the hilltop region. Let us now
choose parameters such that density perturbations with (\ref{COBE}) and
(\ref{7ns}) are realized at $\sigma_* = ( 1 - 10^{-8}) \pi f$. 
Considering the case where $t_{\mathrm{osc}} < t_{\mathrm{reh}}$ and the
curvaton dominates the universe before it decays, 
we adopt the following parameter set (which lies at around the center of
the allowed window in Figure~\ref{fig:ht8dom}):
$H_{\mathrm{inf}} = 10^{6}\, \mathrm{GeV}$, 
$\rho_{\mathrm{reh}}^{1/4} = 10^{9}\, \mathrm{GeV}$
(i.e. $\Gamma_\phi \approx 9.73 \times 10^{-20} M_p$),
$f \approx 1.57 \times 10^{-3} M_p$, and $\Lambda \approx 1.23 \times
10^{-8} M_p$.

\begin{figure}[htbp]
\begin{center}
 \begin{minipage}{.48\linewidth}
  \begin{center}
 \includegraphics[width=\linewidth]{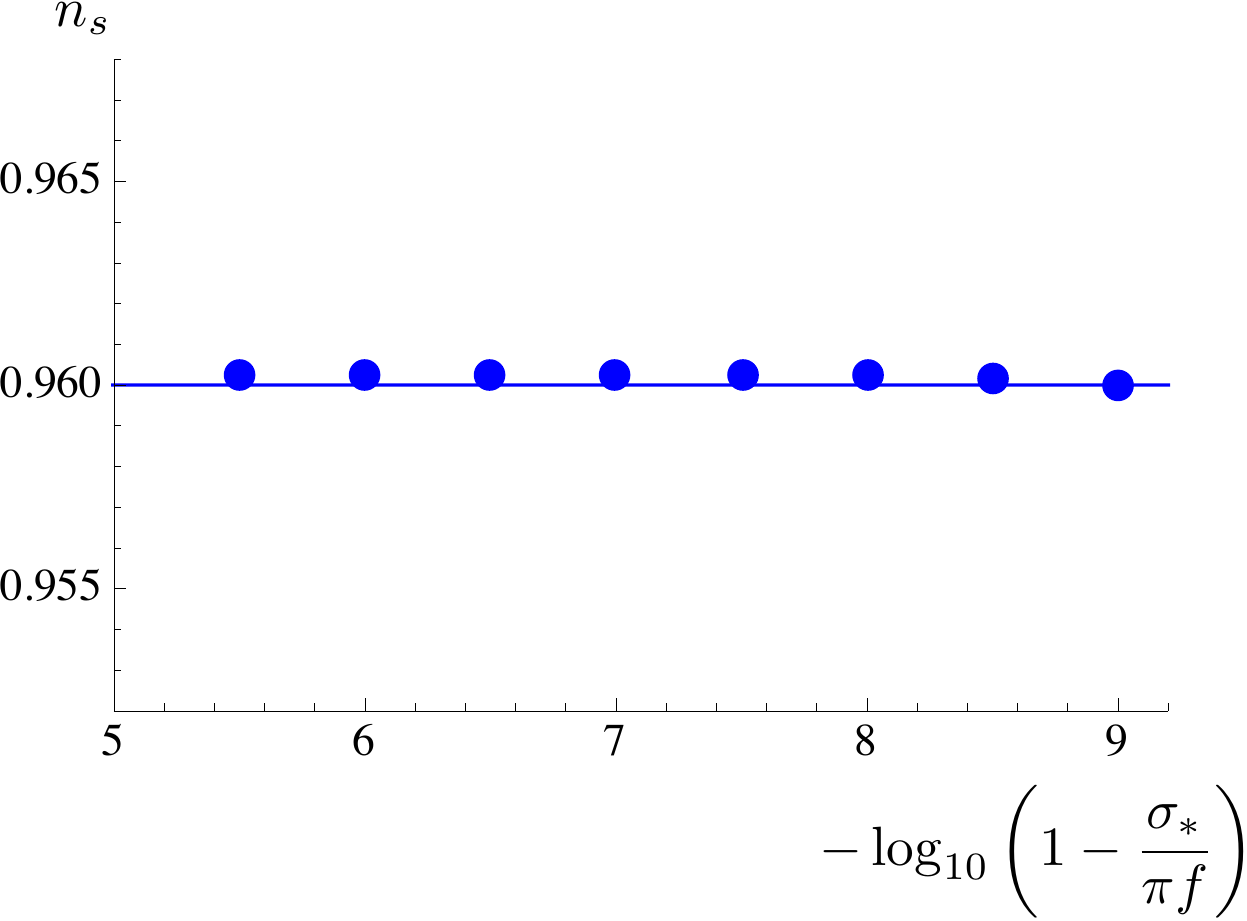}
  \end{center}
  \caption{Spectral index.}
  \label{fig:NGht-ns}
 \end{minipage} 
\end{center}
\end{figure}

\begin{figure}[htbp]
 \begin{minipage}{.48\linewidth}
  \begin{center}
  \includegraphics[width=\linewidth]{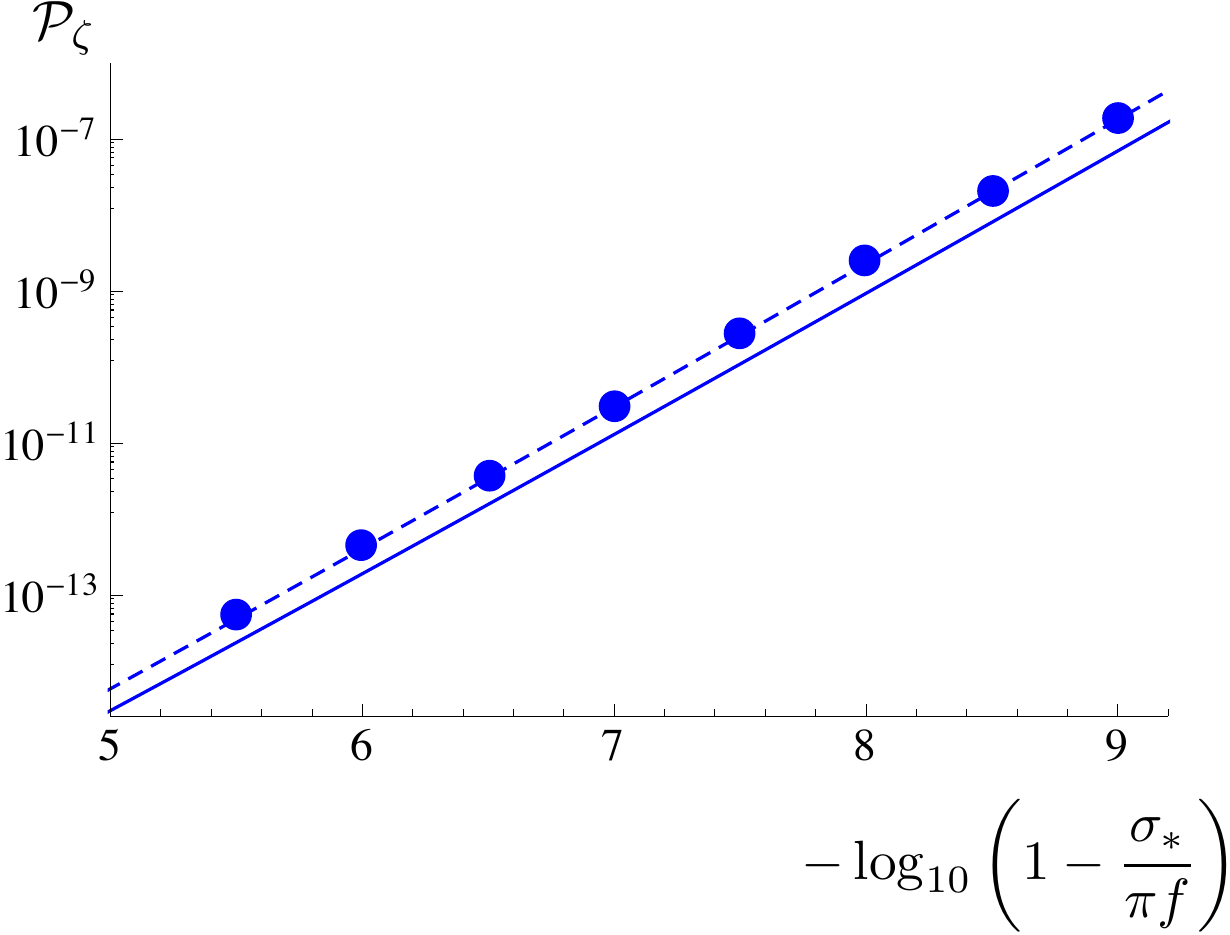}
  \end{center}
  \caption{Power spectrum.}
  \label{fig:NGht-r}
 \end{minipage} 
 \begin{minipage}{0.01\linewidth} 
  \begin{center}
  \end{center}
 \end{minipage} 
 \begin{minipage}{.48\linewidth}
  \begin{center}
 \includegraphics[width=\linewidth]{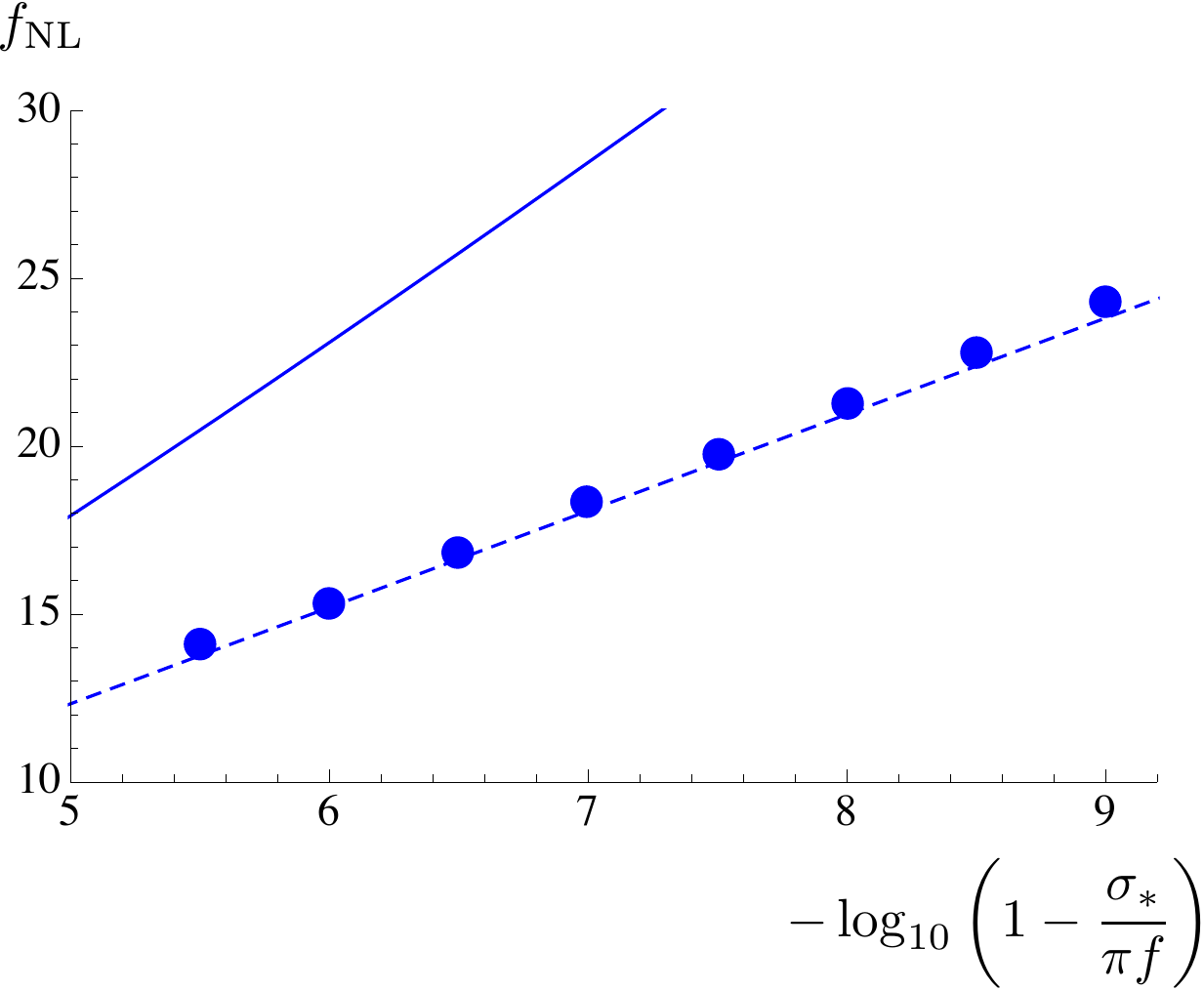}
  \end{center}
  \caption{Non-Gaussianity.}
  \label{fig:NGht-fNL}
 \end{minipage} 
\end{figure}

Introducing $n_*$ as
\begin{equation}
 \alpha_* \equiv 1 - 10^{-n_*},
\end{equation}
the solution of (\ref{4242}) in the region of our interest $2 \lesssim
n_* \lesssim 10$ is approximated by
\begin{equation}
 \alpha_{\mathrm{osc}} \approx 1  - ( 0.416+ 0.0120 n_*) n_*^{-1.47},
 \label{NGhtfit}
\end{equation}
which shows that $\alpha_{\mathrm{osc}}$ approaches unity much
slower than $\alpha_*$ does (cf. Figure~\ref{fig:sigma_osc}). 
Then one can evaluate the density perturbation spectrum as was done for
the non-hilltop region. The results are shown in
Figures~\ref{fig:NGht-ns} - \ref{fig:NGht-fNL}, where again the solid
lines denote the analytic estimations and dots the numerically computed
results. The curvaton well dominates the universe before it decays,
i.e. $r \gg 1$, in the displayed region $5 \lesssim n_* \lesssim 9$. 
Here we should remark that, in the hilltop limit, the
attractor~(\ref{eq17}) (mildly) breaks down before (\ref{dotsigma}) is
realized, hence $\sigma_{\mathrm{osc}}$~(\ref{NGhtfit}) estimated from
(\ref{2424}) contains error of order $\Delta \sigma_{\mathrm{osc}} /
\sigma_{\mathrm{osc}} = \mathcal{O}(1)$.\footnote{The value of $-V' / H
\dot{\sigma}$ (which in the analytic estimations we have treated as a
constant $c=9/2$) becomes as large as $\sim 30$
right before the onset of the oscillation~(\ref{dotsigma}). (This can be
estimated by solving the equation~(\ref{cVpp}) in terms of~$c$.)
Nevertheless, the error in~$\sigma_{\mathrm{osc}}$ is of order unity 
since $-V' / H \dot{\sigma}$ being large is limited to times close
to~$t_{\mathrm{osc}}$.} 
This gives rise to errors of 
the similar order in the analytic estimations, which is clearly seen in
Figure~\ref{fig:NGht-fNL}. 
Therefore, we have also calculated semi-analytic results by
substituting numerically computed $\sigma_{\mathrm{osc}}$ (defined
by (\ref{dotsigma})) into the analytic expressions in
Subsection~\ref{subsec:res}. The semi-analytic results are shown in the
figures as dashed lines, which match well with the numerical
computations. 
(The dashed line is absent in Figure~\ref{fig:NGht-ns} since
the spectral index is evaluated independently
of~$\sigma_{\mathrm{osc}}$). 

The figures show that, as was discussed as generic features of hilltop
curvatons in Subsection~\ref{subsec:hilltop}, the linear order density
perturbations blow up in the hilltop limit, while the non-linearity
parameter~$f_{\mathrm{NL}}$ increases slowly. The spectral index, since
it is determined by the potential curvature, asymptotes to a constant
value as one approaches the hilltop.
However in the extreme hilltop limit the finite size field fluctuations
$\delta \sigma_* = H_{\mathrm{inf}} / 2 \pi$ can no longer be treated as
infinitesimal, giving rise to the tiny variations of the numerical results
seen in Figure~\ref{fig:NGht-ns}.


\subsection{Parameter Space}

Now that we have studied the behaviour of density perturbations from
NG curvatons, let us explore the parameter space which allows generation
of observationally consistent density perturbations. 
Out of the five free parameters ($f$, $\Lambda$,
$H_{\mathrm{inf}}$, $H_{\mathrm{reh}}$, $\sigma_*$), 
two are fixed from the COBE
normalization~(\ref{COBE}) and the WMAP7 central value for the spectral
index~(\ref{7ns}). 
We then seek constraints on the inflation/reheating scales under various
$\sigma_*$, and discuss cosmology with NG curvatons.  

\subsubsection{Requirements}

Let us begin by laying out the requirements for a consistent curvaton
scenario. Firstly, we assume the curvaton energy density to be
negligibly small during inflation,
\begin{equation}
 \frac{V(\sigma_*)}{3 M_p^2 H_{\mathrm{inf}}^2} \ll 1,
 \label{cond3}
\end{equation}
and until the curvaton starts its oscillation
(cf. Assumption~\ref{ass4}),
\begin{equation}
 \frac{V(\sigma_{\mathrm{osc}})}{3 M_p^2 H_{\mathrm{osc}}^2} \ll 1.
 \label{subdomosc}
\end{equation}
Also, quantum fluctuations during inflation should not make the curvaton
jump over its potential minimum in order to avoid the resulting density
perturbations from being highly non-Gaussian, or over the maximum to
avoid domain walls,
\begin{equation}
 \frac{H_{\mathrm{inf}}}{2 \pi } \ll \sigma_*
 \ll \pi f -  \frac{H_{\mathrm{inf}}}{2 \pi }.
\end{equation}
We further require the curvaton's classical rolling to dominate over the
quantum fluctuations during inflation (so that we can have definite
predictions for the perturbation spectrum\footnote{The randomized case
of NG curvatons are discussed in~\cite{Dimopoulos:2003az}.}),
\begin{equation}
  \frac{3}{2 \pi} \frac{H_{\mathrm{inf}}^3}{V'(\sigma_*)} 
  \ll 1,
\end{equation}
where the slow-roll approximation for the curvaton during inflation is
verified by (\ref{cond3}) and (\ref{cond12}). The curvaton decay should
happen after reheating
\begin{equation}
 \Gamma_\phi > \Gamma_\sigma, \label{green}
\end{equation}
and also after the onset of the curvaton oscillation
\begin{equation}
 H_{\mathrm{osc}} > \Gamma_\sigma, \label{4.15}
\end{equation}
but at the latest before the Big Bang Nucleosynthesis (BBN) at temperature
 $T_{\mathrm{BBN}} \sim 1\, \mathrm{MeV}$ \cite{Kawasaki:1999na,Kawasaki:2000en,Hannestad:2004px,Ichikawa:2005vw},
\begin{equation}
 3 M_p^2 \Gamma_\sigma^2 > \left( 1\, \mathrm{MeV} \right)^4.
 \label{BBNconst}
\end{equation}
We also require the oscillating curvaton's mass to be larger than
its decay temperature, in order to avoid possible backreaction effects
to the curvaton's perturbative decay (see
e.g. \cite{Kolb:2003ke,Yokoyama:2005dv,Drewes:2010pf} for discussions on
this issue). Assuming instant thermalization, this condition is written
roughly as 
\begin{equation}
 V''(0) > (3 M_p^2 \Gamma_\sigma^2)^{1/2}.
\end{equation}

The resulting density perturbations should satisfy the COBE normalization,
\begin{equation}
 \mathcal{P}_\zeta  \approx 2.4 \times 10^{-9},
 \label{cond11}
\end{equation}
and for its spectral index we adopt the WMAP7 central value~(\ref{7ns}),
which through (\ref{ns-1}) translates into (note that we are supposing
inflation with $H = \mathrm{const.}$)
\begin{equation}
 \frac{2}{3} \frac{V''(\sigma_*)}{H_{\mathrm{inf}}^2} \approx -0.04. \label{cond12}
\end{equation}
We note that when (\ref{cond12}) is satisfied at $\sigma_* \approx \pi f
/ 2$ with $|\cos (\sigma_*/ f)| \ll 1$, then a rather large
$\Lambda^4 / f^2$ can give rise to a large running of the spectral index
and may contradict with observations. 
However, when fixing \mbox{$n_s - 1 \sim -10^{-2}$}, the allowed
parameter window becomes small anyway as $\sigma_*$ approaches the
inflection point $\pi f /2$ (as we will soon see), 
thus we merely require (\ref{cond12}) for the spectral
tilt. 
Bounds on local-type non-Gaussianity from the WMAP7 (at 95\% CL) are
\begin{equation}
 -10 < f_{\mathrm{NL}} < 74. \label{cond13}
\end{equation}

Constraints on primordial gravitational waves set an upper
bound on the inflationary energy scale. 
The 7-year WMAP+BAO+$H_0$ gives $\mathcal{P}_T / \mathcal{P}_\zeta <
0.24$ (95\% CL), i.e., 
\begin{equation}
 H_{\mathrm{inf}} < 1.3 \times 10^{14}\, \mathrm{GeV}. \label{cond14}
\end{equation}
However, we note that when the inflation scale is high enough to (come
close to) saturate the bound~(\ref{cond14}), then depending on the
inflationary mechanism, one can expect to have a contribution to the
spectral index from a non-vanishing $\dot{H}/H^2$ (see also
Footnote~\ref{foot1}), as well as the central value of the spectral
index bounds~(\ref{7ns}) being slightly shifted (to about 0.97
with the 7-year WMAP+BAO+$H_0$). Here, as we have stated at the
beginning of this subsection, we adopt the value~(\ref{7ns}) and a 
$H=\mathrm{const.}$ inflation, and simply use (\ref{cond14}) as an upper
bound on the inflationary scale. 

Finally, reheating, i.e. inflaton decay, should happen subsequent to inflation,
\begin{equation}
  \Gamma_\phi < H_{\mathrm{inf}}. 
 \label{cond15}
\end{equation}

\subsubsection{Windows for Inflation/Reheating Scales}
\label{subsub:windows}

Among the above constraints, (\ref{cond11}) and (\ref{cond12}) can be used
to fix the parameters $f$ and $\Lambda$. 
Here, note that $\sigma_{\mathrm{osc}}/f$ is estimated as a
function of $\sigma_*/f$ by solving (\ref{4242}) under $\mathcal{N}_* = 50$,
setting $c = 5\,  (\mathrm{or}\, 9/2)$ for $t_{\mathrm{osc}} >
(\mathrm{or}\, <)\, t_{\mathrm{reh}}$, and choosing
$\Lambda^4/f^2 H_{\mathrm{inf}}^2$ as a function of $\sigma_* / f $
from (\ref{cond12}).

\begin{figure}[htbp]
 \begin{minipage}{.48\linewidth}
  \begin{center}
 \includegraphics[width=\linewidth]{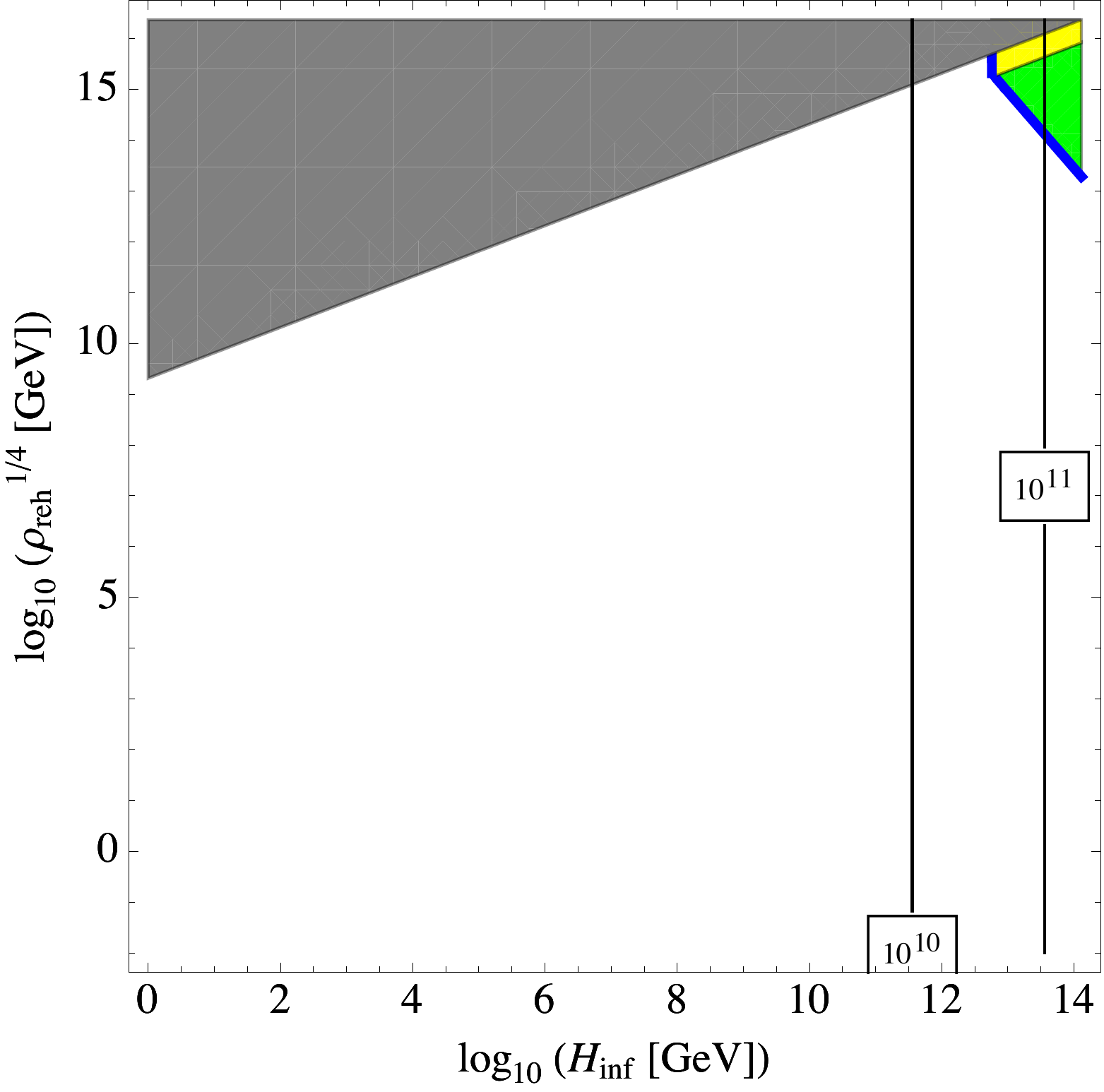}
  \end{center}
  \caption{$\sigma_* / \pi f = 3/4$, $r > 1$.}
  \label{fig:75dom}
 \end{minipage} 
 \begin{minipage}{0.01\linewidth} 
  \begin{center}
  \end{center}
 \end{minipage} 
 \begin{minipage}{.48\linewidth}
  \begin{center}
 \includegraphics[width=\linewidth]{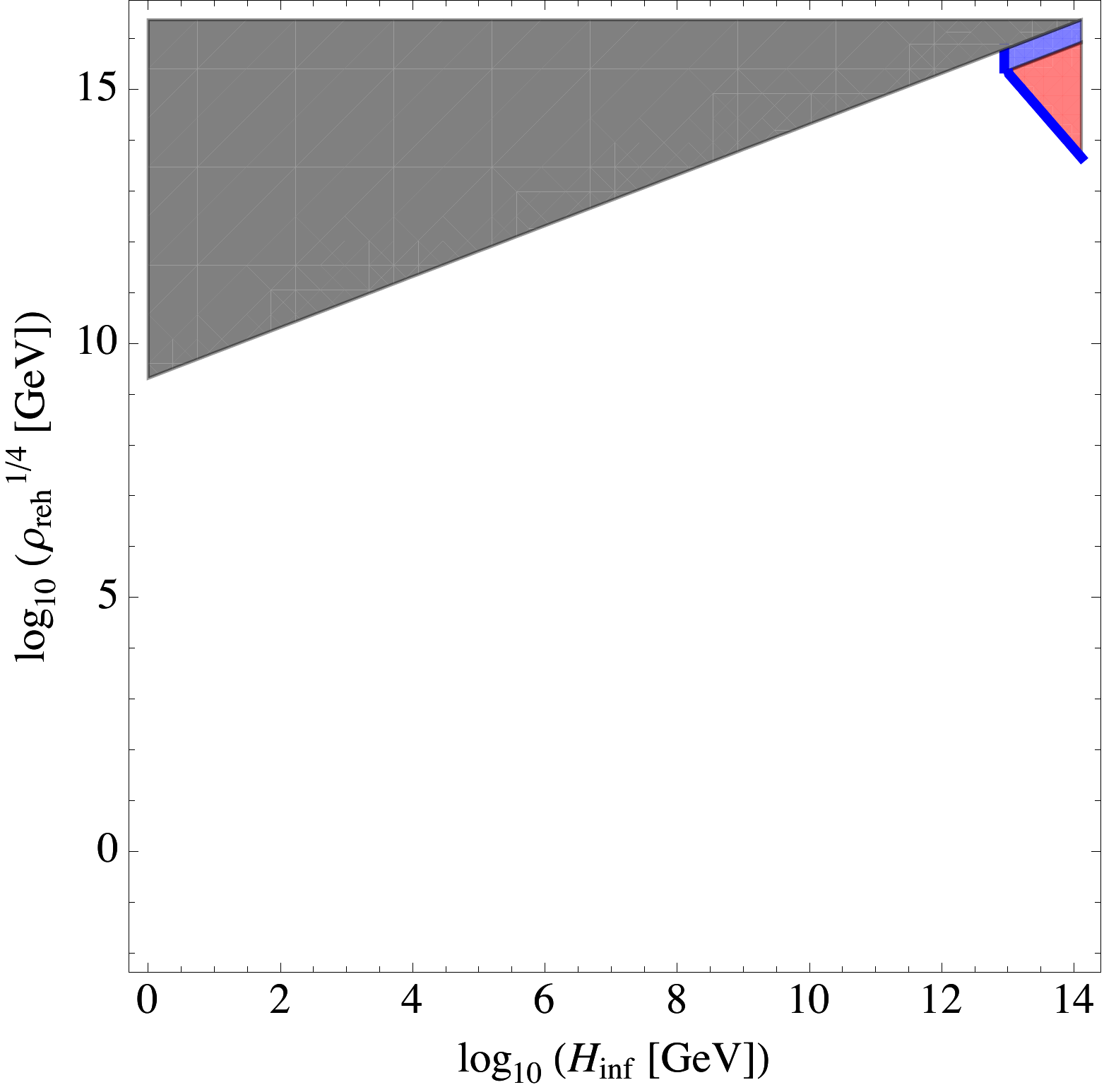}
  \end{center}
  \caption{$\sigma_* / \pi f = 3/4$, $r < 1$.}
  \label{fig:75subdom}
 \end{minipage} 
\end{figure}

\begin{figure}[htbp]
 \begin{minipage}{.48\linewidth}
  \begin{center}
 \includegraphics[width=\linewidth]{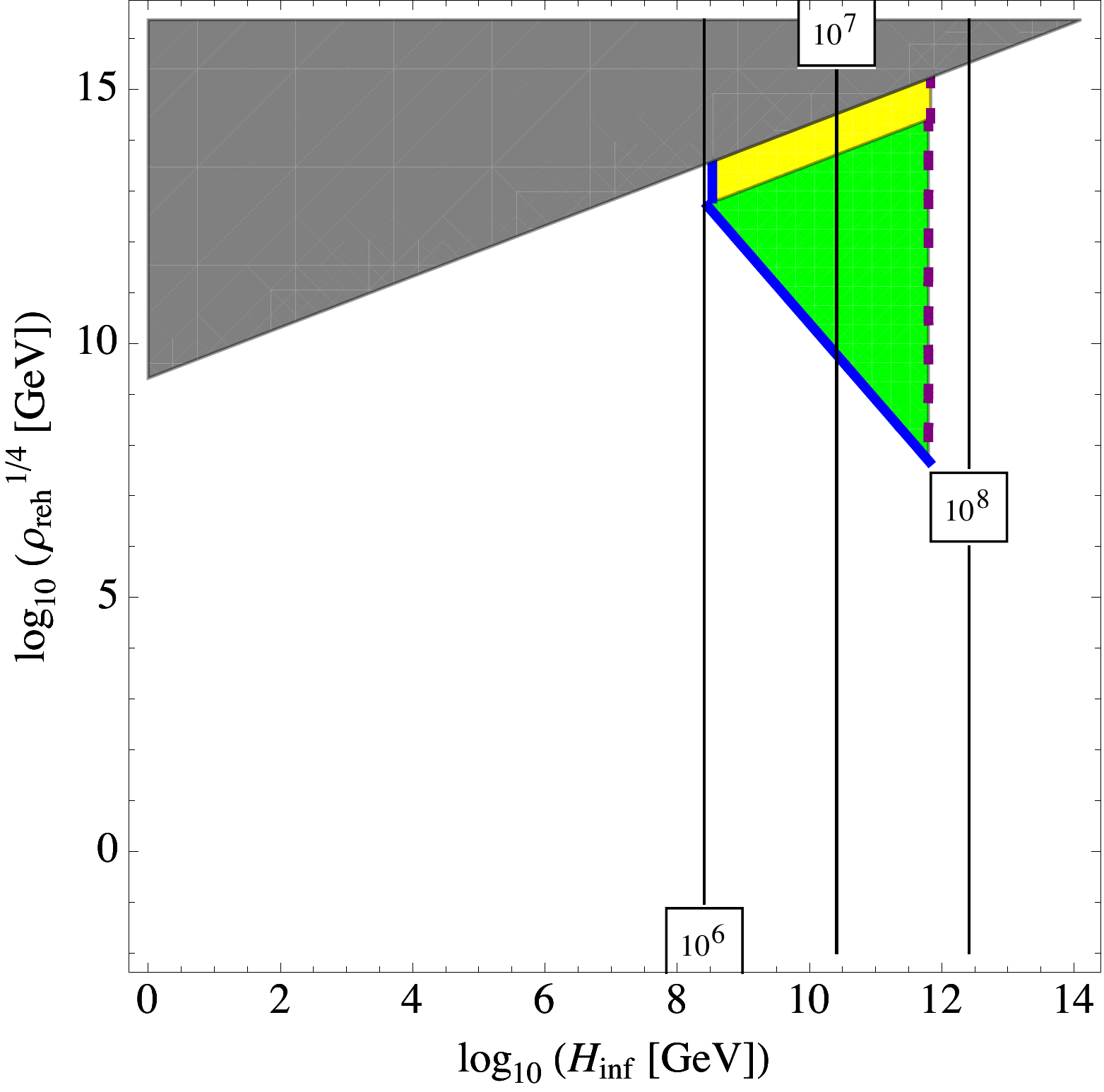}
  \end{center}
  \caption{$\sigma_* / \pi f = 1-10^{-4}$, $r > 1$. \mbox{$f_{\mathrm{NL}}
  \sim 10$} in most of the allowed region.}
  \label{fig:ht4dom}
 \end{minipage} 
 \begin{minipage}{0.01\linewidth} 
  \begin{center}
  \end{center}
 \end{minipage} 
 \begin{minipage}{.48\linewidth}
  \begin{center}
 \includegraphics[width=\linewidth]{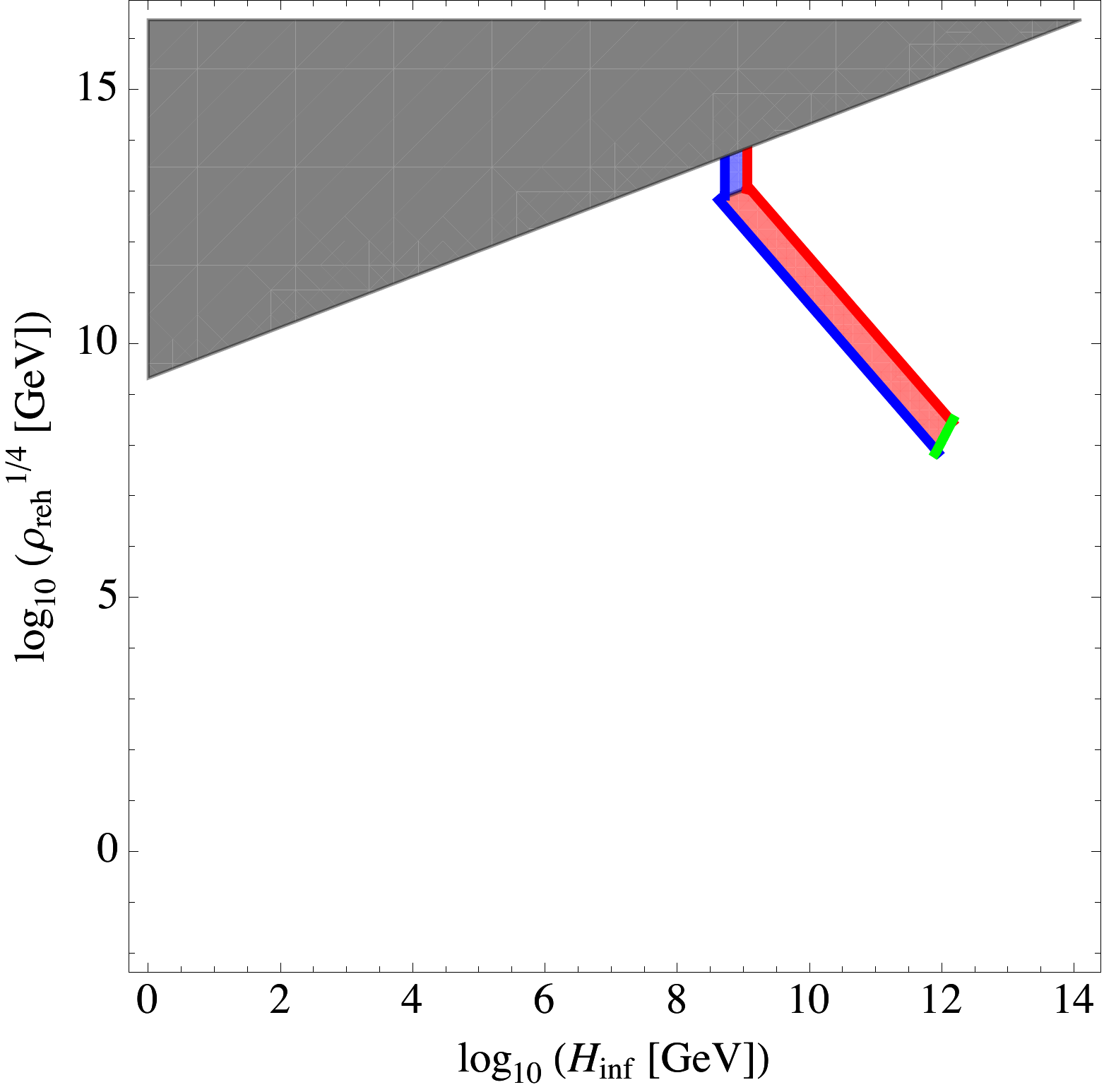}
  \end{center}
  \caption{$\sigma_* / \pi f = 1-10^{-4}$,  $r < 1$.}
  \label{fig:ht4subdom}
 \end{minipage} 
\end{figure}

\begin{figure}[htbp]
 \begin{minipage}{.48\linewidth}
  \begin{center}
 \includegraphics[width=\linewidth]{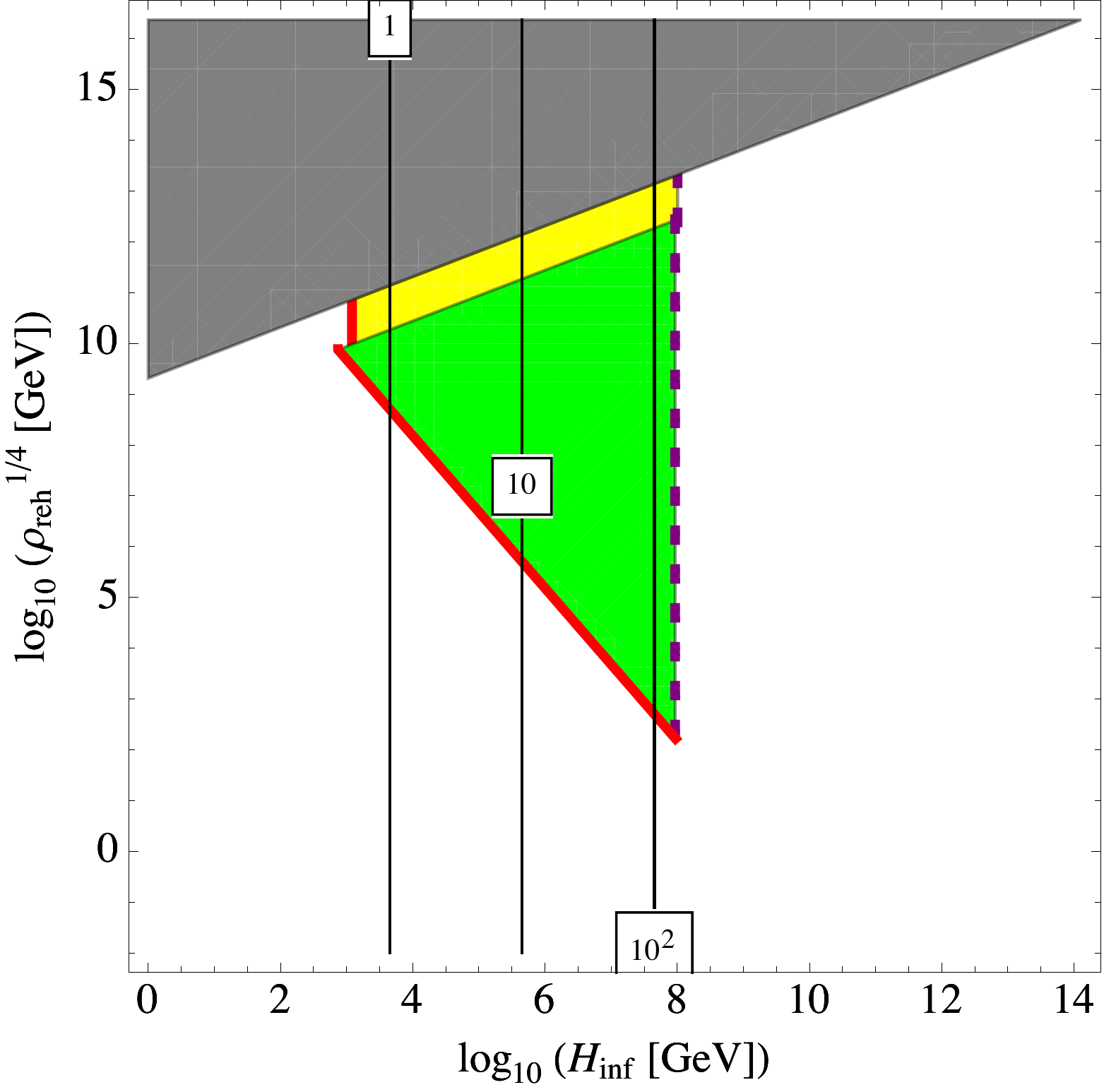}
  \end{center}
  \caption{$\sigma_* / \pi f = 1-10^{-8}$, $r > 1$. \mbox{$f_{\mathrm{NL}}
  \sim 20$} in most of the allowed region.}
  \label{fig:ht8dom}
 \end{minipage} 
 \begin{minipage}{0.01\linewidth} 
  \begin{center}
  \end{center}
 \end{minipage} 
 \begin{minipage}{.48\linewidth}
  \begin{center}
 \includegraphics[width=\linewidth]{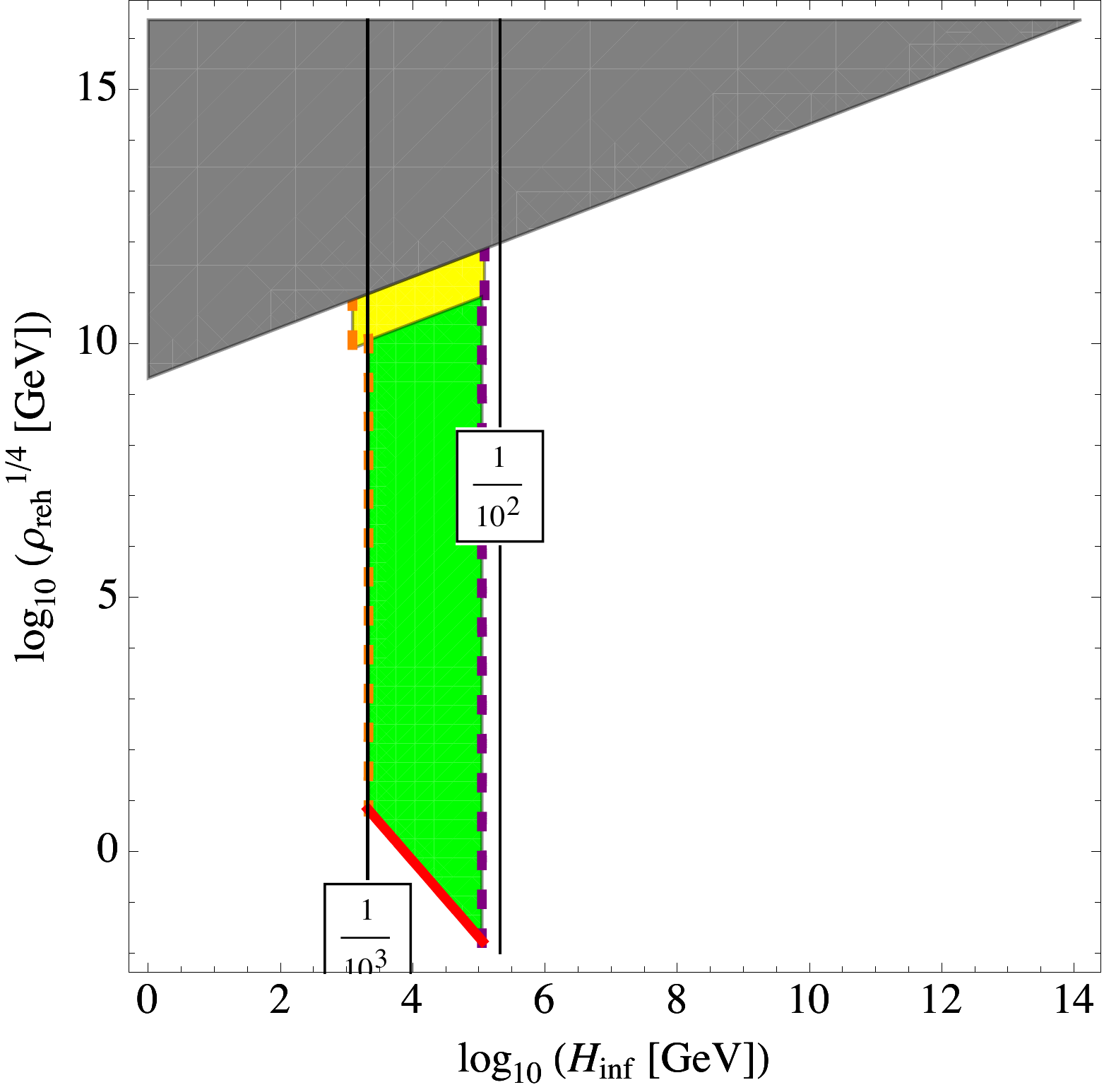}
  \end{center}
  \caption{$\sigma_* / \pi f = 1-10^{-11}$, $r > 1$. \mbox{$f_{\mathrm{NL}}
  \sim 30$} in most of the allowed region.}
  \label{fig:ht11dom}
 \end{minipage} 
\end{figure}

Thus we obtain windows in the parameter space that satisfy all the
above constraints (\ref{cond3}) - (\ref{cond15})
in four regimes, categorized by $r 
\gtrless 1$ and $t_{\mathrm{osc}} \gtrless t_{\mathrm{reh}}$. 
The results are displayed in Figures~\ref{fig:75dom} -
\ref{fig:ht11dom}, in the $H_{\mathrm{inf}}$ -
$\rho_{\mathrm{reh}}^{1/4}$ plane under various values of $\sigma_* / f$. 
The allowed windows are shown as colored regions, where 
the yellow (green) regions correspond to that for
$r > 1$ with $t_{\mathrm{osc}} > t_{\mathrm{reh}}$
($t_{\mathrm{osc}} < t_{\mathrm{reh}}$), 
and the blue (red) regions for $r < 1$ with $t_{\mathrm{osc}} >
t_{\mathrm{reh}}$ ($t_{\mathrm{osc}} < t_{\mathrm{reh}}$),
respectively. 
The right side edge of each box, i.e. the maximum displayed value of
$H_{\mathrm{inf}}$, is set by the gravitational wave
bound~(\ref{cond14}), also the gray shaded region is excluded since
the inflaton should decay after inflation~(\ref{cond15}). 
Differently colored boundaries of the allowed regions
denote different constraints being saturated, where
purple dashed: curvaton being subdominant at the onset of
oscillation~(\ref{subdomosc}),
green: curvaton decay after reheating~(\ref{green}),
orange dashed: BBN~(\ref{BBNconst}), and red: upper bound on
$f_{\mathrm{NL}}$ in~(\ref{cond13}).
Furthermore, the blue lines denote where $r = 1$.
The inequalities~$\ll$ and $\gg$ in the above conditions are relaxed to
$<$ and $>$ upon obtaining the allowed windows in the figures, 
hence in this sense the displayed constraints are conservative
bounds.\footnote{Even under the relaxed conditions, we especially note that 
$\frac{V(\sigma_{\mathrm{osc}})}{3 M_p^2 H_{\mathrm{osc}}^2} < r$ is
guaranteed by the conditions (\ref{green}), (\ref{4.15}), and 
the analytic expression for~$r$~(\ref{anar}).}
Moreover, we note that we have examined the constraints independently in
four distinct regions $t_{\mathrm{osc}} \gg  t_{\mathrm{reh}}$ ($\ll
t_{\mathrm{reh}}$) and $r \gg  1$ ($ \ll 1$), and displayed (two of) them
together in each figure. 
This is why the displayed boundaries bend sharply when crossing 
$t_{\mathrm{osc}} = t_{\mathrm{reh}}$ and moving to differently colored
regions. 
However, we expect that such simplifications and also errors due to
approximations used in the analytic analyses (including that upon
estimating~$\sigma_{\mathrm{osc}}$ in the hilltop limit, see discussions
below (\ref{NGhtfit})) are minor for obtaining order-of-magnitude
estimations of the inflation/reheating scales compatible with
NG curvatons. 
We also show contour lines of the curvaton decay temperature for $r >
1$ cases, where the numbers in boxes denote $ (3 M_p^2
\Gamma_\sigma^2)^{1/4} \, \mathrm{[GeV]}$.
(Note that the contour lines
are actually relevant only on the colored allowed regions.)

First of all, requiring a red-tilted density perturbation spectrum,
the NG curvaton at horizon exit should be located at $1/2 < \alpha_* =
\sigma_* / \pi f < 1$. 
Constraints on NG curvatons away from the hilltop can be seen in 
Figures~\ref{fig:75dom} and \ref{fig:75subdom}, where the representative
value $\alpha_*   = 3/4$ is taken. The dominant and subdominant cases
share the same regions in the \mbox{$H_{\mathrm{inf}}$ -
$\rho_{\mathrm{inf}}^{1/4}$} plane, bounded by (\ref{cond14}) and
(\ref{cond15}). 
Here, one sees that NG curvatons away from the hilltop require rather
high inflation/reheating scales. 
This is because the spectral index $|n_s -1| \sim 10^{-2}$ indicates a
curvaton's effective mass as large as $|m_{\mathrm{eff}}|^2 \sim
10^{-2} H_{\mathrm{inf}}^2$, which starts the curvaton oscillation soon after
inflation ends.\footnote{Strictly speaking, as we have emphasized in the
previous sections, it is not the potential's curvature but rather its
tilt that determines the onset of the oscillation. However, away from
the inflection point $\sigma / \pi f = 1/2$ or the hilltop $\sigma / \pi
f= 1$, the curvature and tilt are of the same scale, i.e. $|V''| \sim V'
/ \sigma$, hence either can be used for estimating~$H_{\mathrm{osc}}$.}
Unless the inflation and reheating scales are high enough to
provide a sufficiently long radiation-dominated era 
for the oscillating curvaton to dominate the universe, the curvaton
cannot source substantial density perturbations. 
This is especially the case as one approaches the inflection point of
the potential, i.e. $\alpha_* \to 1/2$, since then
$|\cos  (\sigma_* / f) | \to 0$ and the tilt of the potential increases,
pushing $H_{\mathrm{osc}}$~(\ref{2424}) as well as the
inflation/reheating scales even higher.

On the other hand, as one approaches the hilltop limit $\alpha_* \to 1$,
the onset of the oscillation is delayed and curvaton domination is
allowed with lower inflation/reheating scales. 
Moreover, the linear order density perturbations obtain an
enhancement factor ~$(1-\alpha_{\mathrm{osc}})/(1-\alpha_*)$, which
should be compensated by a smaller $H_{\mathrm{inf}}/f$
(cf. (\ref{P})). This translates into a larger~$\Lambda$ through the
fixed spectral index (\ref{cond12}), thus further increases the curvaton
energy density fraction, cf.~(\ref{7000}). Figure~\ref{fig:ht4dom}
with $\alpha_*  = 1-10^{-4}$ shows that lower inflation/reheating scales
are compatible with hilltop 
NG curvatons, thus broadening the allowed window. One also sees that
now our requirement of the curvaton being subdominant until the onset of
oscillation (\ref{subdomosc}) becomes important, cutting off the right
edge (i.e. high inflation scales). 

As we have noted in Subsections~\ref{subsec:hilltop} and
\ref{subsubsec:NGht}, non-Gaussianity also increases, though mildly,
towards the hilltop. In the allowed region in Figure~\ref{fig:ht4dom},
the non-Gaussianity is \mbox{$f_{\mathrm{NL}} \sim 10$}, except for
close to the left edge where $r \sim 1$ and hence larger
$f_{\mathrm{NL}}$.
(Here, upon evaluating the explicit value of~$f_{\mathrm{NL}}$, we have
carried out numerical 
calculations in order to avoid errors rising from the approximations
used in the analytic estimations.)
For a subdominant curvaton whose non-Gaussianity is further enhanced,
the observational upper bound on $f_{\mathrm{NL}}$ (\ref{cond13}) becomes a severe
constraint as is seen in Figure~\ref{fig:ht4subdom}.
Eventually the allowed window for subdominant curvatons vanishes as one
approaches the hilltop and the non-Gaussianity increases. 
When $\log_{10} (1-\alpha_*)$ is about $ -8$, the 
window exists only for dominant curvatons,
which is shown in Figure~\ref{fig:ht8dom}. Here, the left edge is now
the non-Gaussianity bound. 

Further approaching the tip, the BBN constraint~(\ref{BBNconst}) kicks
in from low~$H_{\mathrm{inf}}$, cf. Figure~\ref{fig:ht11dom} where
$\alpha_* = 1 - 10^{-11}$.  
The non-Gaussianity in most of the allowed region of
Figure~\ref{fig:ht11dom} (except for close to the bottom edge) is 
$f_{\mathrm{NL}} \sim 30$.
As one goes even closer to the hilltop, the BBN constraint goes towards
higher~$H_{\mathrm{inf}}$ while the constraint~(\ref{subdomosc}) goes
towards lower~$H_{\mathrm{inf}}$, thus removes the allowed window
completely by $\log_{10} (1-\alpha_*) \sim -12$. 

When the curvaton decay rate is further suppressed,
i.e. when $\beta$ in (\ref{beta}) is way smaller than unity, the overall
behaviour of the allowed window stays the same except that the curvaton
decaying temperature is lowered correspondingly.
For instance, when $\beta = 10^{-4}$, 
the blue lines in the figures where $r = 1$ are shifted towards lower 
$H_{\mathrm{inf}}$ by about one order of magnitude 
(or towards lower $\rho_{\mathrm{inf}}^{1/4}$ by about two orders).
The red lines ($f_{\mathrm{NL}}$ bound) shift similarly
while the purple dashed lines (\ref{subdomosc}) are independent
of~$\beta$. Therefore the allowed windows for dominant curvatons slightly
broaden in Figures~\ref{fig:75dom}, \ref{fig:ht4dom}, \ref{fig:ht8dom},
while the windows for subdominant curvatons simply shift towards lower
$H_{\mathrm{inf}}$, $\rho_{\mathrm{inf}}^{1/4}$.
However, since the BBN constraint becomes severer for lower curvaton
decay temperatures, the allowed window for dominant curvatons vanishes at
$\log_{10} (1-\alpha_*) \sim -10$.

Let us comment on implications of the present scenario for the origin of the
baryon asymmetry and dark matter.  Focusing on the case of the dominant curvaton,
the adiabatic density perturbation becomes fixed when the curvaton 
dominates the universe. 
In order to avoid generating too large isocurvature perturbations,
therefore, the baryon as well as dark matter abundance must be fixed after that.
The thermal leptogenesis scenario~\cite{Fukugita:1986hr} requires a relatively
high curvaton decay temperature $T_{\mathrm{dec}} \gtrsim \GEV{9}$. In
the present scenario, $T_{\mathrm{dec}}$
exceeds $\GEV{9}$ for $1- \alpha_* \gtrsim 10^{-3}$.
As $\alpha_*$ further approaches $1$, the decay temperature decreases and
other baryogenesis scenarios are needed. For instance, non-thermal leptogenesis~\cite{Asaka:1999yd}
is known to work  for $T_{\mathrm{dec}} \gtrsim \GEV{6}$, which will be viable for 
$1- \alpha_* \gtrsim 10^{-5}$. For the Affleck-Dine mechanism~\cite{Affleck:1984fy,Dine:1995kz}
to work, the following condition must be met, $H_{\rm dom} > m_{\rm AD}$,
where 
 $H_{\rm dom}$ denotes the Hubble parameter when the curvaton dominates the universe,
 and $m_{\rm AD} (\gtrsim m_{3/2})$ denotes the mass of the Affleck-Dine field.
The above condition can be met for various~$\alpha_*$ as long as the
reheating scale is not so low. 
The electroweak baryogenesis requires the curvaton decay temperature to
exceed of order $ 100$\,GeV, which is translated into $1- \alpha_* \gtrsim 10^{-8}$. 
For $1- \alpha_* \lesssim 10^{-8}$, the decay of $\sigma$ may be able to generate the baryon asymmetry
in a way similar to Ref.~\cite{Cline:1990bw}, using the R-parity violating operators.
On the other hand, there are many dark matter candidates whose abundance is fixed after the curvaton 
dominates the universe. For instance, the (non-)thermal relic of the WIMPs and the QCD axion would
fit with the present  scenario.

\vspace{\baselineskip}

Summarizing, the allowed parameter window for NG curvatons in the 
\mbox{$H_{\mathrm{inf}}$ - $\rho_{\mathrm{inf}}^{1/4}$} plane broadens
significantly in the hilltop regime, especially for $1 - 10^{-4}
\lesssim \sigma_* / \pi f \lesssim 1 - 10^{-11}$. In such hilltop
regime, the NG curvaton mainly dominates the universe before it decays,
but still produces non-Gaussianity in the density perturbations of size
$10 \lesssim f_{\mathrm{NL}} \lesssim 30$.
We reiterate that this prediction on~$f_{\mathrm{NL}}$ is free from the
errors due to approximations used for the analytic estimations, although
the constraints displayed in the figures 
contain errors corresponding to order unity factors.


\subsection{Discussion on Hilltop Curvatons}

Let us comment on initial conditions of hilltop curvaton
models. So far we have simply assumed that the curvaton sits near the
local maximum. 
This may be justified if the curvaton is charged under some symmetry,  which  is unbroken during early stage of the inflation or pre-inflationary epoch.
This ameliorates the amount of fine-tuning of the initial condition for
the hilltop curvaton. On the other hand, in the NG curvaton model,
there is no special point in its field space since the curvaton has an 
approximate shift symmetry. Then, a priori, there is no particular
reason to believe that the curvaton initially sits near the local
maximum. However, we do not know whether the conventional naturalness
argument can be really applied to the determination of parameters in the
early universe. 
For example, the parameters may be determined in a similar fashion as
in the anthropic argument for explaining the present value of the
cosmological constant~\cite{Weinberg:1987dv}.
Suppose that the prior probability of
the initial position of the NG curvaton has a flat distribution in the
field space. Assuming that all the other parameters are fixed, it is
conceivable that only the universe with the curvaton initially sitting
near the local maximum is anthropically selected, if it does not lead to
the habitable universe otherwise. For instance, this is expected to be
the case, if the inflaton mainly decays into the hidden sector while the
curvaton decays into the standard model particles. Thus, the apparent
fine-tuning of the initial condition of the NG curvaton based on the
conventional naturalness argument may be due to our ignorance of the
mechanism about how those parameters are determined.  

Another important constraint on the curvaton model is that it must
account for the red-tilted power spectrum, $n_s - 1 \sim -10^{-2}$. 
As long as the curvaton potential is time-independent, this implies that the curvature of the curvaton potential
 is only about one order of magnitude smaller than the Hubble parameter during inflation. Then, unless the curvaton
 is initially sitting near the hilltop, it starts to oscillate about the potential minimum soon after inflation,
 and the curvaton mechanism does not work for most inflation models. Thus we are led to consider the hilltop curvaton models.
 However, the situation changes if we allow the curvaton potential to depend on time. In particular, the curvaton may have
  a negative Hubble mass during inflation,  $V \supset -  \kappa H^2 \sigma^2$. Then
  the red-tilted spectrum simply implies  $\kappa \sim 10^{-2}$.  After the inflaton decays, the negative Hubble mass term
  disappears, and the curvaton potential may become very flat, and the
curvaton  starts to oscillate long afterward. In this case 
we expect that the curvaton scenario works for many inflation
models while accounting for $n_s -1 \sim -10^{-2}$.\footnote{We note
however that the curvaton dynamics would be 
complicated with a negative Hubble mass term. For instance, the curvaton
slow-rolls on the negative 
Hubble mass away from its origin during inflation, then long after 
reheating, the curvaton comes back and starts to oscillate about the origin. 
Alternatively, the red-tilted power spectrum can be easily realized in the curvaton model with a global U(1) symmetry;
 if the radial component of the complex scalar field moves from a large field value to a smaller one, the phase component (i.e.
 NG curvaton) would have a scale-dependent fluctuation, which results in
 the red-tilted power spectrum. MK and FT thank Shinta Kasuya for
 discussion on this issue.}
 
 Lastly we comment on a possible reason why the NG curvaton has a mass close to the Hubble parameter during inflation.
 If the curvaton potential is generated by non-perturbative dynamics, e.g. strong gauge interactions, the potential
 height is expected to be determined by the dynamical scale, and it is exponentially sensitive to the value of the
 gauge coupling at the UV scale. So, if there is a bias toward a larger value 
of the coupling constant at the UV scale,  the curvaton mass is favored to be as large as possible. 
If the curvaton mass exceeds the Hubble parameter during inflation, it is stabilized at the potential minimum, and
the curvaton mechanism does not work. Thus, the curvaton mass slightly smaller than $H_{\rm inf}$ may be 
selected in the curvaton landscape. Of course this is just a
theoretical possibility, and it is hard to prove that such an argument is
indeed the correct explanation.

\section{Conclusions}
\label{sec:conc}

In this paper, we have explored density perturbations from a curvaton
with a generic potential, and shown that the curvaton's field fluctuations
can lead to non-uniform onset of the curvaton oscillation~$\delta
H_{\mathrm{osc}}$. This sources additional
contributions to the resulting density perturbations, in addition to
that from the field fluctuations directly perturbing the curvaton energy
density. We showed that the combination of the two effects gives
rise to non-trivial behaviours for density perturbations from
curvatons, such as the curvaton sourcing large $f_{\mathrm{NL}}$
with either sign even if it dominates the universe before decay.
The effect due to $\delta H_{\mathrm{osc}}$ becomes important
when the curvaton potential deviates from the familiar quadratic type.
Since a curvaton potential with negative curvature is required during
inflation in order to explain the observationally suggested red-tilted
density perturbation spectrum (without invoking inflation with large
$|\dot{H}/H^2|$), it is of importance to evaluate effects from~$\delta
H_{\mathrm{osc}}$. Our work was also motivated by curvaton models based 
on microscopic physics such as supersymmetry or string theory, 
which provides explicit examples of curvaton potentials possessing
non-trivial forms. 

As a simple case where the $\delta H_{\mathrm{osc}}$ effects become
dominant on the generated density perturbations, we have studied
curvatons that start their oscillations from a flat region along its
potential. Especially for hilltop curvatons, we showed the strong
enhancement of the linear order density
perturbations~$\mathcal{P}_{\zeta}$, accompanied by a 
mild increase of the non-Gaussianity~$f_{\mathrm{NL}}$. 

Another non-trivial case, which we have not studied intensively in this
paper, is when various contributions to the density perturbations cancel
each other and thus suppress the perturbation amplitude. 
Conversely, then the non-Gaussianity is expected to become large. 
It would be interesting to investigate such cases, which may be realized
by, e.g., curvaton potentials possessing plateau regions, or with
superimposed repeated modulations.\footnote{Superimposed repeated
modulations to the inflaton 
potential can also have interesting consequences to the density
perturbation spectrum generated by the inflaton itself, see
e.g.~\cite{Adams:1992bn,Wang:2002hf,Feng:2003mk,Chen:2008wn,Pahud:2008ae,Hamann:2008yx,Flauger:2009ab,Hannestad:2009yx,Flauger:2010ja,Kobayashi:2010pz}.}
We leave this for future work.

We also applied our generic results to study a simple model
where the curvaton is a pseudo-Nambu-Goldstone (NG) boson of a broken U(1)
symmetry. Although NG curvatons away from the hilltop require high
inflation/reheating scales in order to generate 
red-tilted density perturbation spectra with $ n_s - 1 \sim - 10^{-2}$,
it was shown that NG curvatons at the
hilltop can work with a wide range of inflation/reheating scales. We also
showed that working NG curvatons in the hilltop predict the 
non-Gaussianity to lie in the range $10 \lesssim f_{\mathrm{NL}}
\lesssim 30$, which is accessible to upcoming CMB observations such as
the PLANCK satellite.

For the four-point correlation functions of the density perturbations
from hilltop NG curvatons, one obtains relations that are similar
to those for the familiar quadratic curvatons, i.e., $\tau_{\mathrm{NL}}
\sim g_{\mathrm{NL}} \sim f_{\mathrm{NL}}^2$. However, 
different curvaton potentials may lead to different behaviours among the
non-Gaussianities (see related discussions in~\cite{Enqvist:2008gk}). 
It will be interesting to examine systematically the range of
possibilities arising from curvatons with generic energy potentials.

Both from observational and theoretical reasons, it is important to
investigate in detail the conversion process of the curvaton field
fluctuations into the resulting density perturbations. 
In this work, we have provided generic analytic expressions
incorporating effects due to non-uniform onset of the curvaton
oscillation, which open up new possibilities for the curvaton paradigm.

\section*{Acknowledgements}

FT thanks Tomo Takahashi for useful communication on the non-uniform
onset of oscillations in the curvaton scenario.
This work was supported by the Grant-in-Aid for Scientific Research on
Innovative Areas (No. 21111006) [FT,MK], Scientific Research (A)
(No. 22244030 and 21244033 [FT]),  Scientific Research (C) (No. 22540267 [MK]) and
JSPS Grant-in-Aid for Young Scientists (B) (No. 21740160) [FT].  
This work was also supported by World Premier International Center
Initiative (WPI Program), MEXT, Japan.


\appendix

\section{Scalar Field Dynamics in an Expanding Universe}
\label{app:SFD}

In this appendix we discuss dynamics of a scalar field~$\sigma$ rolling
along a potential~$V(\sigma)$ in an expanding universe background. 
Here the potential $V(\sigma)$ is assumed to be a function only
of $\sigma$, and does not depend on say, time, explicitly. 
We especially show a condition under which the equation of motion of the field
\begin{equation}
 \ddot{\sigma} + 3 H(t) \dot{\sigma} + \frac{\partial
  V(\sigma)}{\partial   \sigma} = 0  \label{maru1}
\end{equation}
is approximated by 
\begin{equation}
 c H(t) \dot{\sigma} \simeq - \frac{\partial V(\sigma)}{\partial
  \sigma} , \label{approx}
\end{equation}
where $c$ is a constant.

The background expansion is considered to be independent of the $\sigma$-dynamics
(as in the case of a curvaton before it starts to oscillate), and the
universe to be dominated by some sort of energy density with 
equation of state $p = w \rho$ (with constant~$w$), so
that the Hubble parameter satisfies
\begin{equation}
 \frac{\dot{H}}{H^2} = -\frac{3 (w+1)}{2}.
\end{equation}
By taking a time derivative of both sides of (\ref{approx}) and
comparing with (\ref{maru1}), one can check that
\begin{equation}
 c \simeq 3 - \frac{\dot{H}}{H^2} - \frac{V''}{c H^2}
 = 3 + \frac{3 (w+1)}{2}  - \frac{V''}{c H^2},
 \label{cVpp}
\end{equation}
where a prime denotes a derivative with respect to~$\sigma$. 
Hence for 
\begin{equation}
 \left| \frac{V''}{c H^2} \right| \ll 1, \label{sscond}
\end{equation}
we obtain
\begin{equation}
 c = 3 + \frac{3 (w+1)}{2} \label{thisisc}
\end{equation}
which is no less than 3 for $w \geq -1$. For example, $c =3 $ for a
de~Sitter, $c = 9/2$ for a matter dominated, and $c = 5$ for a radiation 
dominated universe. 
(We note that a non-zero but constant $|V''|/H^2$ can
also realize the approximation~(\ref{approx}) with different values
of~$c$, as in rapid-roll 
inflation~\cite{GarciaBellido:1996ke,Kinney:1997ne,Linde:2001ae,Kinney:2005vj,Tzirakis:2007bf,Kofman:2007tr,Kobayashi:2009nv}.)

One can further show that (\ref{approx}) is an attractor. Introducing
\begin{equation}
 \xi \equiv \frac{\ddot{\sigma}+ 3 H \dot{\sigma}}{c H \dot{\sigma}} - 1
\end{equation}
where we now define $c$ by (\ref{thisisc}), then one can check that
\begin{equation}
 \frac{\dot{\xi}}{H} = -\frac{V''}{c H^2} - c\xi - c \xi^2.
\end{equation}
Since $c$ given by (\ref{thisisc}) is positive, this equation shows that,
at least when $|\xi| < 1$, $|\xi|$ damps as the universe expands until its
amplitude becomes as small as $\mathcal{O}(|V''|/ c^2 H^2)$. Hence as long
as the condition~(\ref{sscond}) holds, the approximation~(\ref{approx})
with (\ref{thisisc}) is a stable attractor.

\section{Density Perturbations from Curvatons with Non-Sinusoidal
 Oscillations} 
\label{app:non-sinu}

In this appendix we generalize the expressions for density perturbations
in the main body of the paper to incorporate a period of
non-sinusoidal curvaton oscillation during which the energy density
redshifts as $\rho_\sigma \propto a^{-n}$ with arbitrary~$n$, followed
by the usual sinusoidal oscillation with $n=3$. The results here can be
applied to, e.g.,  
a curvaton potential which is quadratic around its (local) minimum, but
is dominated by a non-quadratic polynomial term at large~$\sigma$. 

Instead of the assumptions in Subsection~\ref{subsec:deltaN},
here we suppose that at some time~$t_{\mathrm{osc}}$ after inflation the
curvaton starts an oscillation which redshifts away its energy
density as $\rho_\sigma \propto a^{-n}$ with an arbitrary constant~$n$,
until its energy density 
becomes as small as $\rho_\sigma = \rho_{\sigma \mathrm{sin}}$, 
then the curvaton suddenly switches to a sinusoidal oscillation giving
$\rho_\sigma \propto a^{-3}$. After some period of the sinusoidal
oscillation, the curvaton decays to radiation.
Having in mind potentials that switch between quadratic and
non-quadratic polynomials at a certain field value, we take the energy
density~$\rho_{\mathrm{\sigma sin}}$ as a constant which is independent
of $\sigma_*$.  
We also consider the curvaton energy density fraction to be negligibly
small until $t_{\mathrm{reh}}$ or $t_{\mathrm{osc}}$, whichever is
later. 

Then one can compute the $\delta \mathcal{N}$ similarly as in
Subsection~\ref{subsec:deltaN}, yielding
\begin{equation}
 \frac{\partial \mathcal{N}}{ \partial \sigma_*} = 
 \frac{r}{4+3 r} \frac{\partial}{\partial \sigma_*} 
 \left( \frac{3}{n}  \ln \rho_{\sigma \mathrm{osc}} - A \ln
 H_{\mathrm{osc}}^2 \right), 
\label{B1}
\end{equation}
\begin{equation}
 \frac{\partial^2 \mathcal{N}}{ \partial \sigma_*^2} = 
 \frac{16 (1+r)}{ (4+3r) r} 
 \left( \frac{\partial \mathcal{N}}{ \partial \sigma_*}\right)^2
 +  \frac{r}{4+3 r} \frac{\partial^2}{\partial \sigma_*^2} 
 \left(  \frac{3}{n}  \ln \rho_{\sigma \mathrm{osc}} - A \ln
 H_{\mathrm{osc}}^2 \right), 
 \label{B2}
\end{equation}
where $A = 3/4$ for $t_{\mathrm{reh}} < t_{\mathrm{osc}}$, and
$A = 1 $ for $t_{\mathrm{reh}} > t_{\mathrm{osc}}$ (in this latter case 
$t_{\mathrm{reh}}$ can be either before or after $t_{\mathrm{sin}}$,
but of course before $t_{\mathrm{dec}}$).
We note that (\ref{Ar}) holds in the present case as well. It is clear
that the above expressions reproduce those in
Subsection~\ref{subsec:deltaN} for $n = 3$. 
Moreover, the discussions in Subsections~\ref{subsec:CFVOC} and
\ref{sec:H_osc} can be applied to the present case without any modification.

\subsubsection*{example: Linear Curvatons}

As an example for applying the above formulae, let us take a look at a
curvaton potential of the form
\begin{equation}
 V (\sigma) = \Lambda^4 \left(\frac{\sigma}{f}\right)^2 
 \left[ 1 + \left( \frac{\sigma}{f}\right)^2 \right]^{-1/2},
\end{equation}
with constant mass scales $\Lambda$ and $f$. The potential
is quadratic around its minimum $\sigma = 0 $, but approaches a
linear potential at $|\sigma | \gg f$. 

Considering the curvaton to start its oscillation in the linear regime,
i.e. $\sigma_{\mathrm{osc}} \gg f$, then by solving (\ref{sigmaosc}), one finds
\begin{equation}
 \sigma_{\mathrm{osc}} = \frac{2c-6}{2c-5} 
 \left( \sigma_* - \frac{\mathcal{N}_*}{3} \frac{\Lambda^4}{f
  H_{\mathrm{inf}}^2}  \right) .
\end{equation}
The energy density of the curvaton oscillating along the linear regime
redshifts as $\rho_\sigma \propto a^{-2}$, hence one can use the
formulae (\ref{B1}) and (\ref{B2}) to obtain
\begin{equation}
 \frac{\partial \mathcal{N}}{\partial \sigma_*} = 
 \frac{(3+2 A)r}{2 (4+3 r)}
 \left( \sigma_* - \frac{\mathcal{N}_*}{3} \frac{\Lambda^4}{f
  H_{\mathrm{inf}}^2}  \right)^{-1},
\end{equation}
\begin{equation}
 f_{\mathrm{NL}} = \frac{5}{6 r} 
 \left\{ \frac{16 (1+r)}{4+3 r} - \frac{2 (4 +3 r)}{3 + 2 A}  \right\},
\end{equation}
which exhibit behaviours that are rather close to those of curvatons
with quadratic potentials. However, here the spectral index~(\ref{ns-1})
vanishes at leading order, when $\dot{H}_* = 0$.


\clearpage
\begin{thebibliography}{99}


\bibitem{Starobinsky:1980te}
  A.~A.~Starobinsky,
  ``A New Type of Isotropic Cosmological Models Without Singularity,''
  Phys.\ Lett.\  B {\bf 91}, 99 (1980).

\bibitem{Sato:1980yn}
  K.~Sato,
  ``First Order Phase Transition of a Vacuum and Expansion of the Universe,''
  Mon.\ Not.\ Roy.\ Astron.\ Soc.\  {\bf 195}, 467 (1981).

\bibitem{Guth:1980zm}
  A.~H.~Guth,
  ``The Inflationary Universe: A Possible Solution to the Horizon and Flatness
  Problems,''
  Phys.\ Rev.\  D {\bf 23}, 347 (1981).

\bibitem{Mukohyama:2009gg}
  S.~Mukohyama,
  ``Scale-invariant cosmological perturbations from Horava-Lifshitz gravity
  without inflation,''
  JCAP {\bf 0906}, 001 (2009)
  [arXiv:0904.2190 [hep-th]].

\bibitem{Rubakov:2009np}
  V.~A.~Rubakov,
  ``Harrison-Zeldovich spectrum from conformal invariance,''
  JCAP {\bf 0909}, 030 (2009)
  [arXiv:0906.3693 [hep-th]].

\bibitem{Creminelli:2010ba}
  P.~Creminelli, A.~Nicolis and E.~Trincherini,
  ``Galilean Genesis: An Alternative to inflation,''
  JCAP {\bf 1011}, 021 (2010)
  [arXiv:1007.0027 [hep-th]].

\bibitem{Linde:1996gt}
  A.~D.~Linde and V.~F.~Mukhanov,
  ``Nongaussian isocurvature perturbations from inflation,''
  Phys.\ Rev.\  D {\bf 56}, 535 (1997)
  [arXiv:astro-ph/9610219].

\bibitem{Enqvist:2001zp}
  K.~Enqvist and M.~S.~Sloth,
  ``Adiabatic CMB perturbations in pre - big bang string cosmology,''
  Nucl.\ Phys.\  B {\bf 626}, 395 (2002)
  [arXiv:hep-ph/0109214].

\bibitem{Lyth:2001nq}
  D.~H.~Lyth and D.~Wands,
  ``Generating the curvature perturbation without an inflaton,''
  Phys.\ Lett.\  B {\bf 524}, 5 (2002)
  [arXiv:hep-ph/0110002].

\bibitem{Moroi:2001ct}
  T.~Moroi and T.~Takahashi,
  ``Effects of cosmological moduli fields on cosmic microwave background,''
  Phys.\ Lett.\  B {\bf 522}, 215 (2001)
  [Erratum-ibid.\  B {\bf 539}, 303 (2002)]
  [arXiv:hep-ph/0110096].

\bibitem{Enqvist:2005pg}
  K.~Enqvist, S.~Nurmi,
  JCAP {\bf 0510}, 013 (2005).
  [astro-ph/0508573].

\bibitem{Enqvist:2008gk}
  K.~Enqvist and T.~Takahashi,
  ``Signatures of Non-Gaussianity in the Curvaton Model,''
  JCAP {\bf 0809}, 012 (2008)
  [arXiv:0807.3069 [astro-ph]].

\bibitem{Enqvist:2009zf}
  K.~Enqvist, S.~Nurmi, G.~Rigopoulos, O.~Taanila and T.~Takahashi,
  ``The Subdominant Curvaton,''
  JCAP {\bf 0911}, 003 (2009)
  [arXiv:0906.3126 [astro-ph.CO]].

\bibitem{Kawasaki:2008mc}
  M.~Kawasaki, K.~Nakayama, F.~Takahashi,
  ``Hilltop Non-Gaussianity,''
  JCAP {\bf 0901}, 026 (2009).
  [arXiv:0810.1585 [hep-ph]].

\bibitem{Sasaki:2006kq}
  M.~Sasaki, J.~Valiviita and D.~Wands,
  ``Non-Gaussianity of the primordial perturbation in the curvaton model,''
  Phys.\ Rev.\  D {\bf 74}, 103003 (2006)
  [arXiv:astro-ph/0607627].

\bibitem{Larson:2010gs}
  D.~Larson {\it et al.},
  ``Seven-Year Wilkinson Microwave Anisotropy Probe (WMAP) Observations: Power
  Spectra and WMAP-Derived Parameters,''
  Astrophys.\ J.\ Suppl.\  {\bf 192}, 16 (2011)
  [arXiv:1001.4635 [astro-ph.CO]].

\bibitem{Dunkley:2010ge}
  J.~Dunkley {\it et al.},
  ``The Atacama Cosmology Telescope: Cosmological Parameters from the 2008
  Power Spectra,''
  arXiv:1009.0866 [astro-ph.CO].

\bibitem{Hlozek:2011pc}
  R.~Hlozek {\it et al.},
  ``The Atacama Cosmology Telescope: a measurement of the primordial power
  spectrum,''
  arXiv:1105.4887 [astro-ph.CO].

\bibitem{Lyth:1996im}
  D.~H.~Lyth,
  ``What would we learn by detecting a gravitational wave signal in the cosmic
  microwave background anisotropy?,''
  Phys.\ Rev.\ Lett.\  {\bf 78}, 1861 (1997)
  [arXiv:hep-ph/9606387].

\bibitem{Hamaguchi:2003dc}
  K.~Hamaguchi, M.~Kawasaki, T.~Moroi, F.~Takahashi,
  ``Curvatons in supersymmetric models,''
  Phys.\ Rev.\  {\bf D69}, 063504 (2004).
  [hep-ph/0308174].

\bibitem{Kobayashi:2009cm}
  T.~Kobayashi, S.~Mukohyama,
  ``Curvatons in Warped Throats,''
  JCAP {\bf 0907}, 032 (2009).
  [arXiv:0905.2835 [hep-th]].

\bibitem{Burgess:2010bz}
  C.~P.~Burgess, M.~Cicoli, M.~Gomez-Reino, F.~Quevedo, G.~Tasinato and I.~Zavala,
  ``Non-standard primordial fluctuations and nongaussianity in string
  inflation,''
  JHEP {\bf 1008}, 045 (2010)
  [arXiv:1005.4840 [hep-th]].

\bibitem{Dimopoulos:2003az}
  K.~Dimopoulos, D.~H.~Lyth, A.~Notari, A.~Riotto,
  ``The Curvaton as a pseudoNambu-Goldstone boson,''
  JHEP {\bf 0307}, 053 (2003).
  [hep-ph/0304050].

\bibitem{Dimopoulos:2003ss}
  K.~Dimopoulos, G.~Lazarides, D.~Lyth and R.~Ruiz de Austri,
  ``Curvaton dynamics,''
  Phys.\ Rev.\  D {\bf 68}, 123515 (2003)
  [arXiv:hep-ph/0308015].

\bibitem{Starobinsky:1986fxa}
  A.~A.~Starobinsky,
  ``Multicomponent de Sitter (Inflationary) Stages and the Generation of
  Perturbations,''
  JETP Lett.\  {\bf 42}, 152 (1985)
  [Pisma Zh.\ Eksp.\ Teor.\ Fiz.\  {\bf 42}, 124 (1985)].

\bibitem{Sasaki:1995aw}
  M.~Sasaki and E.~D.~Stewart,
  ``A General Analytic Formula For The Spectral Index Of The Density
  Perturbations Produced During Inflation,''
  Prog.\ Theor.\ Phys.\  {\bf 95}, 71 (1996)
  [arXiv:astro-ph/9507001].

\bibitem{Wands:2000dp}
  D.~Wands, K.~A.~Malik, D.~H.~Lyth and A.~R.~Liddle,
  ``A new approach to the evolution of cosmological perturbations on large
  scales,''
  Phys.\ Rev.\  D {\bf 62}, 043527 (2000)
  [arXiv:astro-ph/0003278].

\bibitem{Lyth:2004gb}
  D.~H.~Lyth, K.~A.~Malik and M.~Sasaki,
  ``A general proof of the conservation of the curvature perturbation,''
  JCAP {\bf 0505}, 004 (2005)
  [arXiv:astro-ph/0411220].

\bibitem{Komatsu:2001rj}
  E.~Komatsu and D.~N.~Spergel,
  ``Acoustic signatures in the primary microwave background bispectrum,''
  Phys.\ Rev.\  D {\bf 63}, 063002 (2001)
  [arXiv:astro-ph/0005036].

\bibitem{Kawasaki:1999na}
  M.~Kawasaki, K.~Kohri, N.~Sugiyama,
  ``Cosmological constraints on late time entropy production,''
  Phys.\ Rev.\ Lett.\  {\bf 82}, 4168 (1999).
  [astro-ph/9811437].

\bibitem{Kawasaki:2000en}
  M.~Kawasaki, K.~Kohri, N.~Sugiyama,
  ``MeV scale reheating temperature and thermalization of neutrino background,''
  Phys.\ Rev.\  {\bf D62}, 023506 (2000).
  [astro-ph/0002127].

\bibitem{Hannestad:2004px}
 S.~Hannestad,
 ``What is the lowest possible reheating temperature?,''
 Phys.\ Rev.\  D {\bf 70}, 043506 (2004).

\bibitem{Ichikawa:2005vw}
 K.~Ichikawa, M.~Kawasaki and F.~Takahashi,
 ``The oscillation effects on thermalization of the neutrinos in the  universe
 with low reheating temperature,''
 Phys.\ Rev.\  D {\bf 72}, 043522 (2005).

\bibitem{Kolb:2003ke}
  E.~W.~Kolb, A.~Notari and A.~Riotto,
  ``On the reheating stage after inflation,''
  Phys.\ Rev.\  D {\bf 68}, 123505 (2003)
  [arXiv:hep-ph/0307241].

\bibitem{Yokoyama:2005dv}
  J.~Yokoyama,
  ``Can oscillating scalar fields decay into particles with a large thermal
  mass?,''
  Phys.\ Lett.\  B {\bf 635}, 66 (2006)
  [arXiv:hep-ph/0510091].

\bibitem{Drewes:2010pf}
  M.~Drewes,
  ``On the Role of Quasiparticles and thermal Masses in Nonequilibrium
  Processes in a Plasma,''
  arXiv:1012.5380 [hep-th].

\bibitem{Fukugita:1986hr}
  M.~Fukugita, T.~Yanagida,
  ``Baryogenesis Without Grand Unification,''
  Phys.\ Lett.\  {\bf B174}, 45 (1986);
  For a review, see W.~Buchmuller, R.~D.~Peccei, T.~Yanagida,
  ``Leptogenesis as the origin of matter,''
  Ann.\ Rev.\ Nucl.\ Part.\ Sci.\  {\bf 55}, 311-355 (2005).
  [hep-ph/0502169].  

\bibitem{Asaka:1999yd}
  T.~Asaka, K.~Hamaguchi, M.~Kawasaki and T.~Yanagida,
  ``Leptogenesis in inflaton decay,''
  Phys.\ Lett.\ B {\bf 464}, 12 (1999)
  [arXiv:hep-ph/9906366];
  ``Leptogenesis in inflationary universe,''
  Phys.\ Rev.\ D {\bf 61}, 083512 (2000)
  [arXiv:hep-ph/9907559].
  
\bibitem{Affleck:1984fy}
  I.~Affleck and M.~Dine,
  ``A New Mechanism For Baryogenesis,''
  Nucl.\ Phys.\  B {\bf 249}, 361 (1985).

\bibitem{Dine:1995kz}
  M.~Dine, L.~Randall and S.~D.~Thomas,
  ``Baryogenesis From Flat Directions Of The Supersymmetric Standard Model,''
  Nucl.\ Phys.\  B {\bf 458}, 291 (1996)
  [arXiv:hep-ph/9507453].

\bibitem{Cline:1990bw}
  J.~M.~Cline and S.~Raby,
  ``Gravitino induced baryogenesis: A Problem made a virtue,''
  Phys.\ Rev.\  D {\bf 43}, 1781 (1991).

\bibitem{Weinberg:1987dv}
  S.~Weinberg,
  ``Anthropic Bound on the Cosmological Constant,''
  Phys.\ Rev.\ Lett.\  {\bf 59}, 2607 (1987).

\bibitem{Adams:1992bn}
  F.~C.~Adams, J.~R.~Bond, K.~Freese, J.~A.~Frieman and A.~V.~Olinto,
  ``Natural inflation: Particle physics models, power law spectra for large
  scale structure, and constraints from COBE,''
  Phys.\ Rev.\  D {\bf 47}, 426 (1993)
  [arXiv:hep-ph/9207245].

\bibitem{Wang:2002hf}
  X.~Wang, B.~Feng, M.~Li, X.~L.~Chen and X.~Zhang,
  ``Natural inflation, Planck scale physics and oscillating primordial
  spectrum,''
  Int.\ J.\ Mod.\ Phys.\  D {\bf 14}, 1347 (2005)
  [arXiv:astro-ph/0209242].

\bibitem{Feng:2003mk}
  B.~Feng, M.~z.~Li, R.~J.~Zhang and X.~m.~Zhang,
  ``An inflation model with large variations in spectral index,''
  Phys.\ Rev.\  D {\bf 68}, 103511 (2003)
  [arXiv:astro-ph/0302479].

\bibitem{Chen:2008wn}
  X.~Chen, R.~Easther and E.~A.~Lim,
  ``Generation and Characterization of Large Non-Gaussianities in Single Field
  Inflation,''
  JCAP {\bf 0804}, 010 (2008)
  [arXiv:0801.3295 [astro-ph]].

\bibitem{Pahud:2008ae}
  C.~Pahud, M.~Kamionkowski and A.~R.~Liddle,
  ``Oscillations in the inflaton potential?,''
  Phys.\ Rev.\  D {\bf 79}, 083503 (2009)
  [arXiv:0807.0322 [astro-ph]].

\bibitem{Hamann:2008yx}
  J.~Hamann, S.~Hannestad, M.~S.~Sloth and Y.~Y.~Y.~Wong,
  ``Observing trans-Planckian ripples in the primordial power spectrum with
  future large scale structure probes,''
  JCAP {\bf 0809}, 015 (2008)
  [arXiv:0807.4528 [astro-ph]].

\bibitem{Flauger:2009ab}
  R.~Flauger, L.~McAllister, E.~Pajer, A.~Westphal and G.~Xu,
  ``Oscillations in the CMB from Axion Monodromy Inflation,''
  JCAP {\bf 1006}, 009 (2010)
  [arXiv:0907.2916 [hep-th]].

\bibitem{Hannestad:2009yx}
  S.~Hannestad, T.~Haugbolle, P.~R.~Jarnhus and M.~S.~Sloth,
  ``Non-Gaussianity from Axion Monodromy Inflation,''
  JCAP {\bf 1006}, 001 (2010)
  [arXiv:0912.3527 [hep-ph]].

\bibitem{Flauger:2010ja}
  R.~Flauger and E.~Pajer,
  ``Resonant Non-Gaussianity,''
  JCAP {\bf 1101}, 017 (2011)
  [arXiv:1002.0833 [hep-th]].

\bibitem{Kobayashi:2010pz}
  T.~Kobayashi and F.~Takahashi,
  ``Running Spectral Index from Inflation with Modulations,''
  JCAP {\bf 1101}, 026 (2011)
  [arXiv:1011.3988 [astro-ph.CO]].

\bibitem{GarciaBellido:1996ke}
  J.~Garcia-Bellido, D.~Wands,
  ``The Spectrum of curvature perturbations from hybrid inflation,''
  Phys.\ Rev.\  {\bf D54}, 7181-7185 (1996).
  [astro-ph/9606047].

\bibitem{Kinney:1997ne}
  W.~H.~Kinney,
  ``A Hamilton-Jacobi approach to nonslow roll inflation,''
  Phys.\ Rev.\  {\bf D56}, 2002-2009 (1997).
  [hep-ph/9702427].

\bibitem{Linde:2001ae}
  A.~D.~Linde,
  ``Fast roll inflation,''
  JHEP {\bf 0111}, 052 (2001).
  [hep-th/0110195].

\bibitem{Kinney:2005vj}
  W.~H.~Kinney,
  ``Horizon crossing and inflation with large eta,''
  Phys.\ Rev.\  {\bf D72}, 023515 (2005).
  [gr-qc/0503017].

\bibitem{Tzirakis:2007bf}
  K.~Tzirakis, W.~H.~Kinney,
  ``Inflation over the hill,''
  Phys.\ Rev.\  {\bf D75}, 123510 (2007).
  [astro-ph/0701432].

\bibitem{Kofman:2007tr}
  L.~Kofman, S.~Mukohyama,
  ``Rapid roll Inflation with Conformal Coupling,''
  Phys.\ Rev.\  {\bf D77}, 043519 (2008).
  [arXiv:0709.1952 [hep-th]].

\bibitem{Kobayashi:2009nv}
  T.~Kobayashi, S.~Mukohyama, B.~A.~Powell,
  ``Cosmological Constraints on Rapid Roll Inflation,''
  JCAP {\bf 0909}, 023 (2009).
  [arXiv:0905.1752 [astro-ph.CO]].


\end{thebibliography}
\end{document}